\documentclass[twocolumn,apj]{emulateapj}
\usepackage{color}

\begin{document}

\title{Continuum Enhancements in the Ultraviolet, the Visible and the Infrared during the X1 flare on 2014 March 29}

\author{Lucia Kleint\altaffilmark{1}, Petr Heinzel\altaffilmark{2}, Phil Judge\altaffilmark{3}, S\"am Krucker\altaffilmark{1,4}}
\altaffiltext{1}{University of Applied Sciences and Arts Northwestern Switzerland, Bahnhofstrasse 6, 5210 Windisch, Switzerland}
\altaffiltext{2}{Astronomical Institute, The Czech Academy of Sciences, Fri\v{c}ova 298, 25165 Ond\v{r}ejov, Czech Republic}
\altaffiltext{3}{NCAR/HAO, P.\,O.\,Box 3000, Boulder CO 80307, USA}
\altaffiltext{4}{Space Sciences Laboratory, University of California, Berkeley, CA, USA}
%\affil{$^1$LMSAL/BAER Institute, USA; kleintl@ucar.edu

%\and

\begin{abstract}
Enhanced continuum brightness is observed in many flares (``white light flares''), yet it is still unclear which processes contribute to the emission. To understand the transport of energy needed to account for this emission, we must first identify both the emission processes and the emission source regions.
Possibilities include heating in the chromosphere causing optically thin or thick emission from free-bound transitions of Hydrogen, and heating of the photosphere causing enhanced H$^-$ continuum brightness. To investigate these possibilities, we combine observations from IRIS, SDO/HMI, and the ground-based FIRS
instrument, covering wavelengths in the far-UV, near-UV, visible, and infrared during the X1 flare SOL20140329T17:48. Fits of blackbody spectra to infrared and visible wavelengths are reasonable, yielding radiation temperatures $\sim$6000-6300 K. The NUV emission, formed in the upper photosphere under undisturbed conditions, exceeds these simple fits during the flare, requiring extra emission from the Balmer continuum in the chromosphere. Thus, the continuum originates from enhanced radiation from photosphere (visible-IR) and chromosphere (NUV). From the standard thick-target flare model, we
calculate the energy of the nonthermal electrons observed by RHESSI and compare it to the energy radiated by the continuum emission. We find that the energy contained in most electrons $>$40 keV, or alternatively, of $\sim$10-20\% of electrons $>$20 keV is sufficient to explain the extra continuum emission of $\sim4-8 \times 10^{10}$ erg s$^{-1}$ cm$^{-2}$. Also, from the timing of the RHESSI
HXR and the IRIS observations, we conclude that the NUV continuum is emitted nearly instantaneously when HXR emission is observed with a time difference of no more than 15 s.
 \end{abstract}
\keywords{Sun: flares --- Sun: continuum}

\section{Introduction}

\subsection{The problem of the origin of White Light Emission}
The origin of white light (WL) emission during flares is still debated. First discovered by \citet{carrington1859} as a bright visible emission during a large flare, it is now known to be common, even for relatively weak flares \citep{matthewsetal2003,hudsonetal2006,jessetal2008}, and can be easily found in SDO/HMI data \citep{juanetal2011}. But the question remains as to which mechanisms contribute to this strong increase of radiation throughout the whole spectrum.

This question is important to resolve for at least two reasons. Firstly, the observed energy radiated away is a large fraction of flare energy that is unavailable to generate dynamic pulses that sometimes propagate into the solar interior, generating ``sunquakes''.  Secondly, in the standard flare model, the flare energy is released in the overlying corona, but the mode(s) by which energy propagates downwards to produce the enhanced continuum
  radiation from the dense photosphere or chromosphere is a subject of debate.  WL emission has even been observed at 1.56$\mu$ \citep{xuetal2004}. At this wavelength, the photospheric opacity is a minimum, and under undisturbed conditions, the emerging radiation forms a few dozen km below the 500 nm continuum photosphere.  Overall, we must consider transport of energy across some 15 scale heights spanning conditions in the corona to those in the solar interior!

\subsection{A conceptual guide to formation of continua in flares}

Below we will discuss the formation of broad-band continua, at visible, UV and IR wavelengths, during flares.  We will use non-LTE radiative transfer models. Most of the recent flare literature uses considerably simpler physical pictures \citep[e.g.,][]{kerrfletcher2014}, involving formation of the spectrum in the photosphere and/or chromosphere. To clarify and connect these pictures we note that at wavelength $\lambda$, with source function $S_\lambda(\tau_\lambda)$ (units erg~cm$^{-2}$~s$^{-1}$~sr$^{-1}$~\AA$^{-1}$) 
the intensity emerging from a semi-infinite atmosphere observed  at optical depth $\tau_\lambda=0$ and $\mu=1$ is
\begin{equation}
I_\lambda = \int_0^\infty S_\lambda(\tau_\lambda) \exp(-\tau_\lambda) d \tau_\lambda
\end{equation}
For formation under optically thick conditions, the Eddington-Barbier solution ($S=a+b\tau$)
 to the above is simply
\begin{equation}
I_\lambda \approx S_\lambda(\tau_\lambda=1) \label{eq:onesource}
\end{equation}
Thus, for a photospheric-like temperature structure (source function decreasing with 
height), the observed intensity is merely the source function where 
the optical depth is unity.  In photospheric layers, LTE is often reasonable especially for continua, 
and then the source function is the Planck function $B_\lambda$, so that
\begin{equation}
I_\lambda \approx B_\lambda(\tau_\lambda=1). \label{eq:onesourceb}
\end{equation}
When light at wavelengths of interest forms at a similar depth and thus temperature (i.e., 
$\tau_\lambda \approx {\rm constant}$), a Black Body (BB) spectrum will arise.  This will motivate our use of BB 
spectra below. If however we have a homogeneous optically thin ($\tau_\lambda= \delta$) emission layer with source function
$S'$ above the emitting 
photosphere, the approximate intensity from these arguments is seen to be 
\begin{eqnarray}
I_\lambda &\approx& B_\lambda(\tau_\lambda=1) (1 - \delta)   +  S'_\lambda \delta\\
         &\approx& B_\lambda(\tau_\lambda=1) (1 - \delta)   +  \epsilon' dz,  \label{eq:twosources}
\end{eqnarray}
where $S'=\epsilon'/\kappa'$ is the chromospheric source function,
$\epsilon'$ is the emission coefficient and $\kappa'$ the opacity of the chromosphere (in cm$^{-1}$).  Using these equations we can thereby examine spectrum formation in the optically thick photosphere or
chromosphere (equation \ref{eq:onesource} and \ref{eq:onesourceb}). In the case where an optically thinner chromosphere contributes emission to
an underlying photosphere (equation \ref {eq:twosources}), the last term of the equation corresponds to the observed continuum enhancement.

\subsection{The problem addressed in the present paper}

  In the thick target model \citep{brown1971}, accelerated electrons leading to hard X-ray ($\sim$30-100 keV)
  Bremsstrahlung emission cannot penetrate into the photosphere, yet WL
  sources, also seen in photospheric spectral lines, often correspond
  to HXR sources in shape and location
  \citep[e.g.][]{kruckeretal2011}.  Several pictures of energy
  transport and spectrum formation 
have been proposed to reconcile these curious observations
  \citep[see
    e.g.,][]{cramwoods1982,machadoemslieavrett1989,kerrfletcher2014}:

\begin{itemize}

\item{} The continuum enhancements (``CEs'', sometimes also ``continuum excess'') arise mostly from deposition of energy
  in the \textit{chromosphere}, for example by electron or ion beams. The hydrogen in the chromosphere is strongly ionized by collisions with both thermal (increased temperature and density) as well as non-thermal (beams) electrons and starts to recombine. The H$^-$ opacity is negligible in the chromosphere.
   In this
  case, the broad-band continuum opacities are dominated by the
  bound-free transitions of hydrogen with edges at 911, 3646, 8203
  \AA. WL continua can be thin or thick through the chromosphere,
  depending on the nature of the heating, which determines the
  populations of the $n=2,3\ldots$ levels in hydrogen, meaning that significant heating increases the opacity and optical thickness. In the optically thin case, the CE will be proportional to $\kappa_{\lambda} S_\lambda d$, $d$ being the thickness of the layer and $\kappa_{\lambda}$ the opacity (from Eq.~\ref{eq:twosources}).
  
 item{} The CEs arise mostly from deposition of  energy in the \textit{photosphere}, probably not by electrons,  because they are assumed to be stopped in higher layers. In this case the WL continua are dominated by a combination of H$^-$ and H bound-free opacity, except in the UV
 where other elements can contribute. Possible mechanisms include Alfv\'en waves \citep[e.g.]{fletcherhudson2008} and fast protons. However, the former have been discounted for the flare under scrutiny here \citep{judgeetal2014} and the latter are extremely difficult to observe. In rare proton observations, their location does not  always coincide with the HXR signature of electrons  \citep{hurfordetal2003}. 

\item{} Energy dumped in the chromosphere might then penetrate
  downwards to heat the photosphere radiatively \textit{(``radiative
    backwarming''}).  However, this energy always must compete with
  the flux density of $6.27\times10^{10}$ erg~cm$^{-2}$~s$^{-1}$
  emerging from the solar interior.  Backwarming is expected to be important
  only near or below the temperature minimum region, where sufficient
  H$^-$ population densities and continuum opacity exists.  The
  emitted spectrum will be a combination of the reprocessed downward
  radiation, a BB-like spectrum from H$^-$, plus Balmer and Paschen
  continuum emission below the edges.

\end{itemize}

In all cases the emergent spectra should show Balmer and perhaps
Paschen jumps, when the $H^-$ emission processes do not dominate.
Between 911 and $\approx 2000$\AA, photoionization of ``metals''
dominates the opacity and the spectra form exclusively in the
chromosphere.  The near-UV (NUV) continuum is difficult to identify in a
forest of lines, its enhancement was first detected by
\citet{heinzelkleint2014} with IRIS data.

In this paper, we uniquely combine simultaneous data from four widely spaced
spectral windows: the far-UV (FUV; at 1330 \AA, IRIS), NUV (at 2830 \AA,
IRIS), visible (at 6173~\AA, HMI), and IR (10840~\AA, FIRS) obtained
during the X1 flare SOL20140329T17:48. Our goal is to constrain the
processes that contribute to continuum emission, to compare its energy
to energy deposited by accelerated electrons, and to investigate the
temporal relation between HXR emission and NUV continuum emission.
Given the diverse conditions under which FUV-NUV-visble and IR spectra
form, we take care when applying elementary concepts found in earlier
literature such as ``black body spectra'' and ``optically thin
recombination spectra'' to our flare data.

%http://adsabs.harvard.edu/abs/1980ApJ...239L..27M

%http://adsabs.harvard.edu/abs/1985SoPh...98..255B %boyer et al

%temp of blackbody decreases to 8000 during gradual phase. balmer evolves more slowly and blackbody and persists during gradual phase, height of balmer jump increases.

%################################################### ################################################################
\section{Observations and data reduction}
\label{obs}

%We use IRIS and RHESSI to investigate UV continuum and HXR emission during 
The X1 flare on 2014-03-29 \citep{kleintetal2015} 
occurred in AR 12017 at heliocentric angle $\mu=0.82$ with its (GOES)
maximum at 17:48 UT. We use data from IRIS (UV), RHESSI (X-ray), SDO/HMI (visible) and FIRS (IR). Other instruments 
on Hinode and the IBIS instrument at the Dunn Solar telescope 
were operated in modes leading to no useful additional continuum data.
%For Hinode, this is due to the lack of continuum (the Ca K and \ion{Na}{1} D1 lines were observed), or the timing of the SP raster (slit was already too far west when the flare started). For IBIS, the prefilters are too narrow to include a continuum (in H$\alpha$ and \ion{Ca}{2} 8542), and the broadband ``continuum'' data showed a higher variation due to seeing than due to the WL emission with much lower resulting relative continuum enhancements than HMI.
Spectral imaging data from Hinode lacked continuum channels (only the Ca K and \ion{Na}{1} D1 lines were observed), and the SP slit was too far west when the flare started. For IBIS, the prefilters are too narrow to include a continuum (in H$\alpha$ and \ion{Ca}{2} 8542), and the broadband ``continuum'' data showed a higher variation due to seeing than due to the WL emission with much lower resulting relative CEs than HMI.

\subsection{IRIS}

The Interface Region Imaging Spectrograph \citep[IRIS,][]{iris2014} records spectra in the FUV and NUV, plus simultaneous slitjaw images (SJI).
We use the same IRIS X1-flare data as in \citet{heinzelkleint2014}. In summary, they are 8-step rasters with a cadence of 75 s and a FOV of 14\arcsec$\times$174\arcsec\ for the FUV and NUV spectra and 174\arcsec$\times$174\arcsec\ for the SJI. The spatial resolution was 0\farcs166/px and the spectral resolution 25.46 m\AA\ pixel$^{-1}$.

While all FUV spectra had an exposure time of 8 s, the NUV spectra were exposure controlled, reducing their exposure time from 8 s to 2.4 s during the flare. This is not obvious using the default routine \textit{read\_iris\_l2.pro}, and therefore, we additionally manually corrected the NUV headers for the exposure time, and for the different resulting \textit{date\_obs}. This correction is important for an accurate comparison to RHESSI, where an offset of $\sim$6 s already may be visible in the reconstructed images. The IRIS data were aligned to AIA 1600 data, which included a rotation of 0.5 degrees and a small (few arcsec) shift. The final alignment is estimated to be better than 0\farcs3.

\subsubsection{IRIS Absolute Calibration}
  We convert the measured counts per second into absolute units (erg s$^{-1}$ cm$^{-2}$ sr$^{-1}$ \AA$^{-1}$) by applying the calibration developed by J. P. Wuelser and H. Tian. It converts the measured intensity $I_m$ ([DN/s]) into absolute intensity $I_{abs}$ by
  \begin{equation} \label{eq:cal}
  I_{abs} = I_m \ x \ \frac{hc}{\lambda} \frac{1}{A_{\rm eff}\ d \ \omega}.
  \end{equation}
 $x$ is the number of photons per DN (18 for NUV spectra, 4 for FUV spectra). The energy is calculated with the Planck constant $h=6.63 \times 10^{-27}$ erg s, the speed of light $c=3\times$10$^{10}$ cm s$^{-1}$, and the wavelength $\lambda$ [cm]. The normalization is done with the effective area $A_{\rm eff}$, which was measured pre-launch and estimated post-launch and is available in SolarSoft through \textit{iris\_get\_response()}, the dispersion $d=0.0254$ \AA\ pixel$^{-1}$, and the solid angle $\omega$, calculated as slit width $\times$ pixel size along slit, $\omega$ = 0\farcs33 $\times$ 0\farcs166 $\times$ (725 km/arcsec)$^2$ /  ($1.496 \times 10^8$ km)$^2$. We used both pre- and post-launch estimations of the effective area.

 According to J. P. Wuelser (private communication), the FUV effective area varies significantly compared to pre-launch values, while the NUV is more constant. We therefore decreased the pre-launch effective area of FUV to 38\% of its value according to an estimate valid for the end of March 2014. For the NUV, the effective area may have increased by about 15 \% by March 2014, compared to pre-launch values. The post-launch values provided by \textit{iris\_get\_response()} already take these degradations into account automatically. We note however, that it is unclear which values are more accurate for our observing date, because the post-launch calibration resulted from a cross-calibration with SORCE/SOLSTICE, which occurred about half a year after our flare.

\subsection{RHESSI}
Data from the \textit{Reuven Ramaty High Energy Solar Spectroscope Imager} \citep[RHESSI,][]{linetal2002} were used to trace the locations of accelerated electrons through HXR emission. CLEAN images were reconstructed for each exact IRIS observing time $t_{\rm IRIS}$ with an integration  time of $t_{\rm IRIS} \pm 6$s in the energy range of 30--100~keV. 
The biggest issue combining multiple instruments in the analysis is their co-alignment. 
%After carefully study, we derived an empirical roll correction for RHESSI 
%by comparing it to the ribbons visible in several AIA wavelengths and the WL emission in HMI. 
The default reconstructed RHESSI images were offset from ribbon features in AIA by about 3\arcsec. Hinode, which matched AIA perfectly (which however may have resulted from the automatic data reduction) indicated that the offset between RHESSI and AIA/Hinode was spurious, as it would be unphysical to have both flare footpoints in the same magnetic polarity and having HXR emission trail behind the ribbon emission. A closer inspection of RHESSI roll angle data (PMTRAS) revealed that the star field, which is used to find RHESSI's roll orientation, was sparse with only one usable star at the time around flare maximum. While this is not conclusive evidence which instrument may be offset, it strongly suggests that there may be a difference in the roll angles of the instruments, as those are less precisely known than the x-y offsets. Therefore, all our RHESSI data, unless otherwise specified in the text, were rotated by an empirical roll correction of 0.2$^\circ$ clockwise about Sun center.

 \begin{figure*} % fig 1
  \centering 
   \includegraphics[width=.57\textwidth]{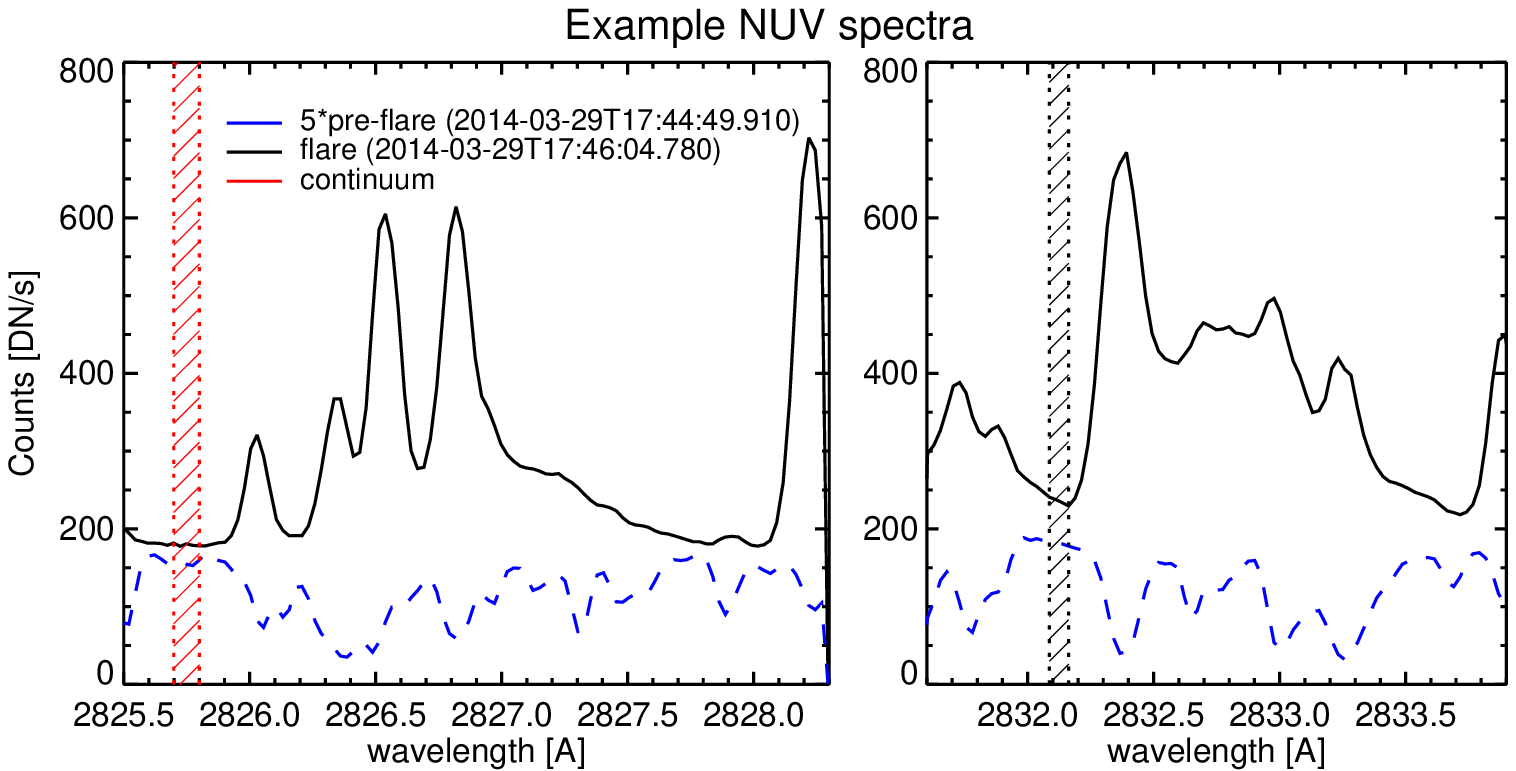}
   \includegraphics[width=.42\textwidth]{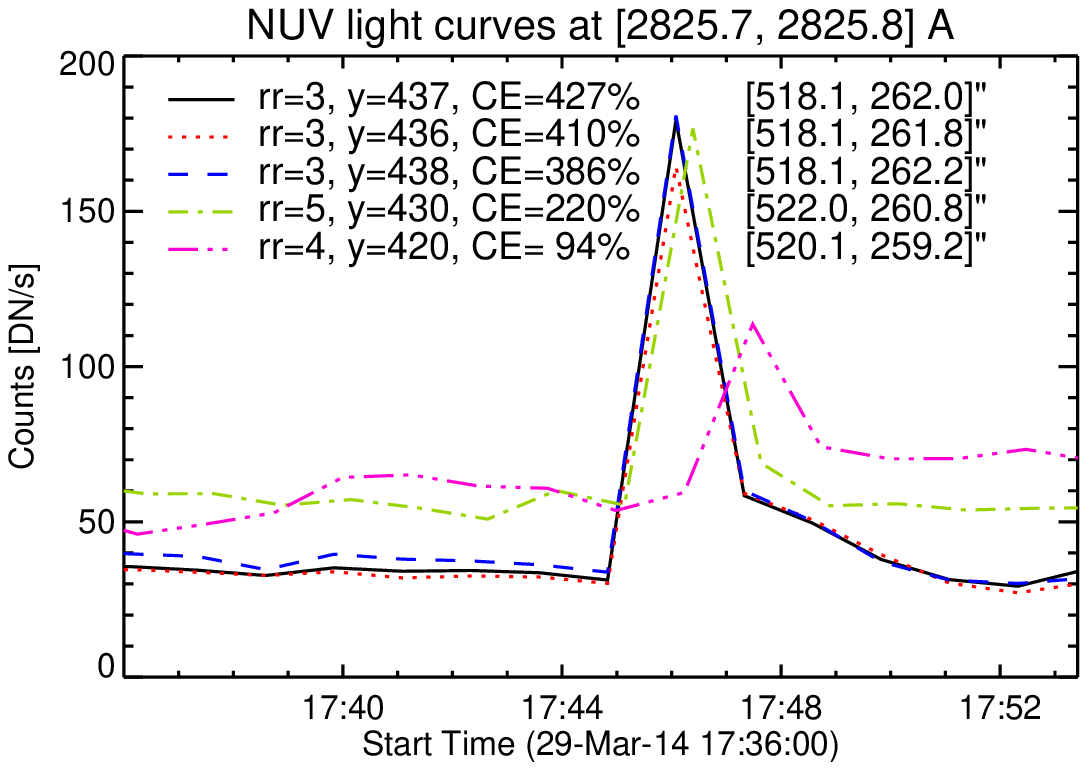}
   \includegraphics[width=.45\textwidth, bb=0 -2 340 224,clip]{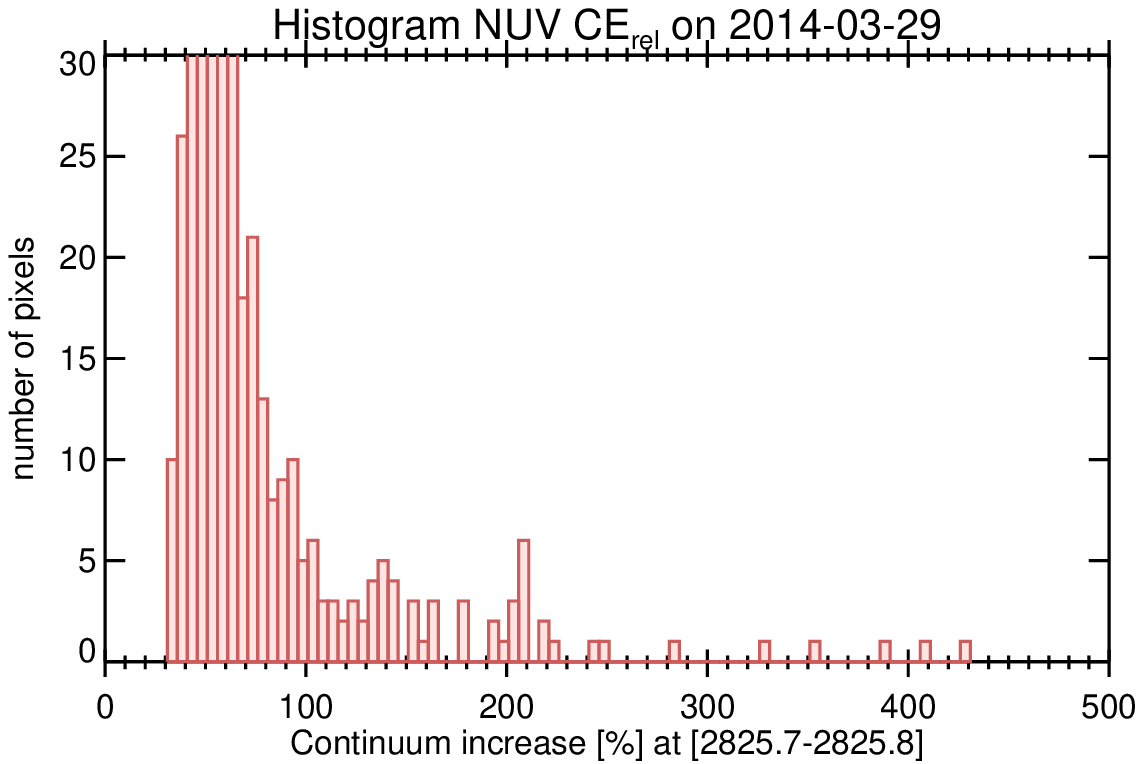}
      \includegraphics[width=.51\textwidth,bb=0 7 453 283,clip]{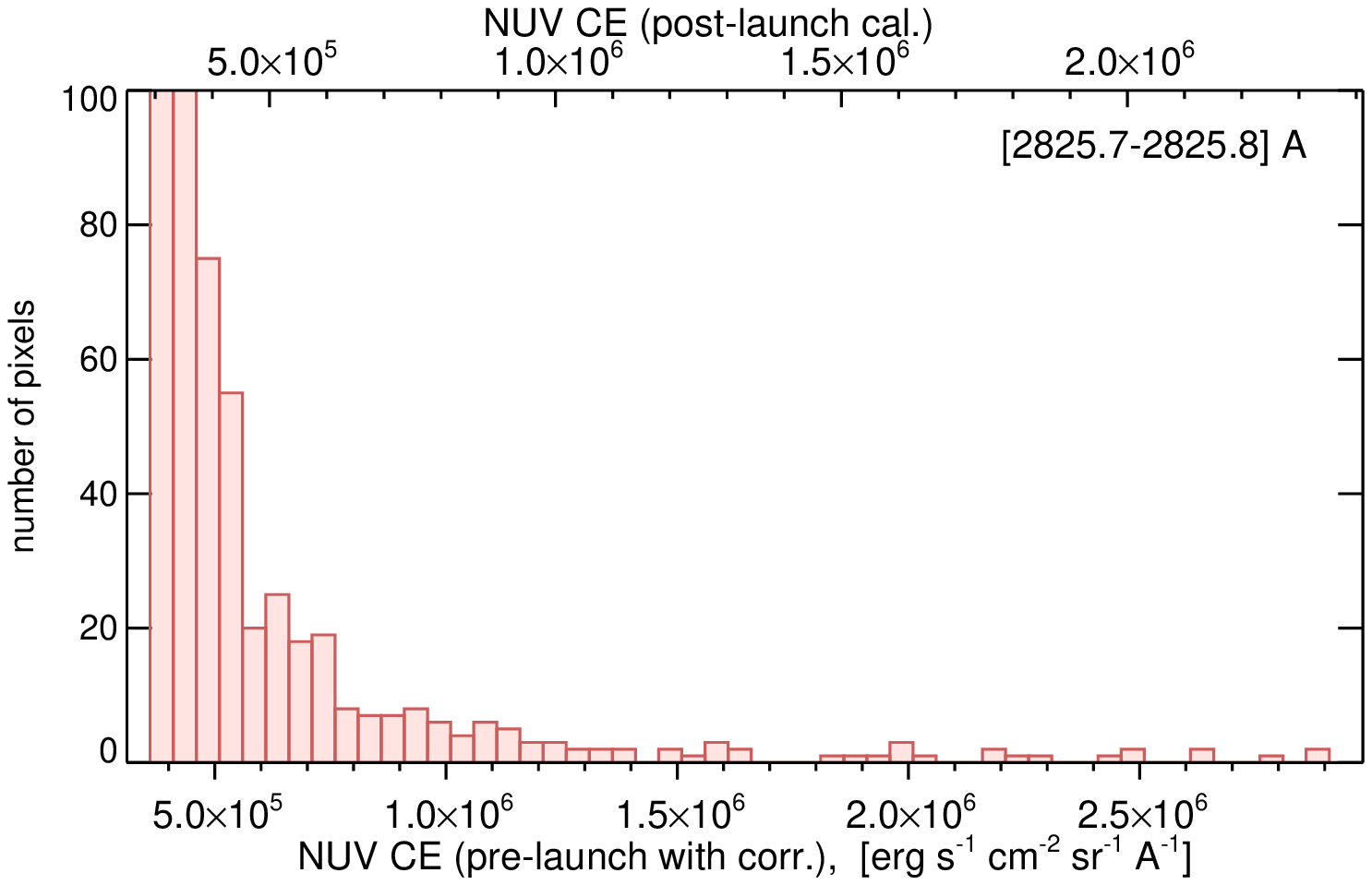}
   \caption{Top row: Left panel: Example NUV pre-flare (blue dashed, scaled by a factor of 5) and flare (black solid) spectra of the pixel that showed maximum enhancement. Our chosen quasi-continuum is marked with red stripes, another possible quasi-continuum, which showed larger variations, is marked in the dark striped region. Right panel: Example light curves of the NUV continuum, with their origin indicated (rr=raster step, starting at 0, y=pixel along slit, solar coordinates given on the right). Bottom row: 
   Histogram of the relative CEs from IRIS NUV data during the X1-flare on 2014-03-29. Most pixels show values below 200\%, but there are few large enhancements up to $\sim$430\%. The right panel shows  NUV CEs (=excess) in absolute radiometric units for post-launch and pre-launch calibrations.}
        \label{ex}
  \end{figure*}

\subsection{HMI}

Data from the Helioseismic and Magnetic Imager  \citep[HMI,][]{scherrerhmi2012} contain a "quasi-continuum" near the \ion{Fe}{1} 6173 \AA\ line. Because the JSOC series \textit{hmi.Ic\_45s} is interpolated over five temporal intervals with a sinc function \citep{juanetal2011}, this may create spurious signals during the rapidly changing conditions during flares. We therefore use \textit{hmi.Ic\_45s\_nrt} data, which are linearly interpolated over only three temporal intervals. Apart from a rotation by the value from HMI's CROTA2 header keyword, no other alignment was performed. The HMI plate scale is 0\farcs504 pixel$^{-1}$ and the FWHM of the filter passband is 76 m\AA.

 \begin{figure*} % fig 2
  \centering 
   \includegraphics[width=.43\textwidth]{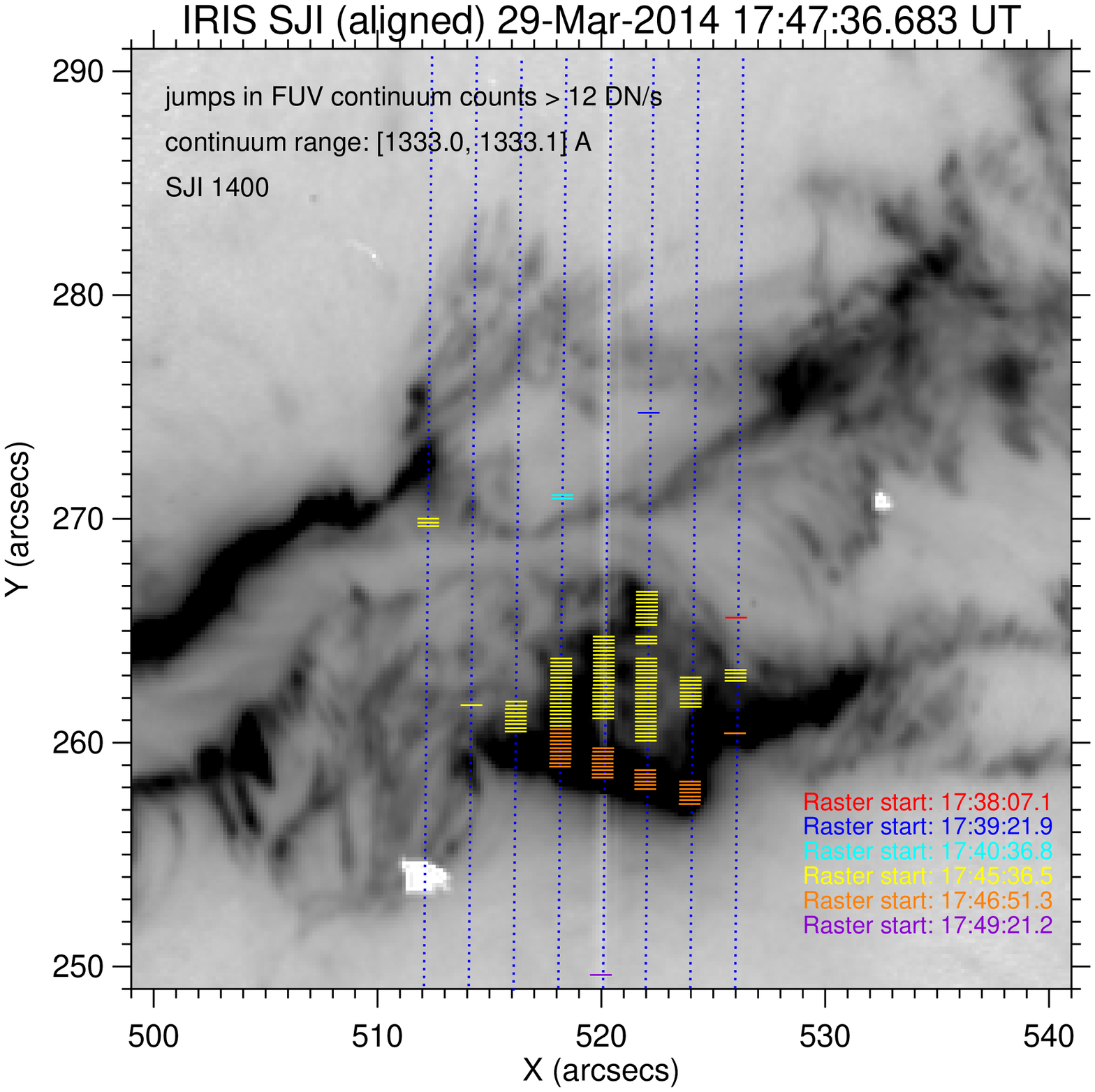}
     \includegraphics[width=.47\textwidth]{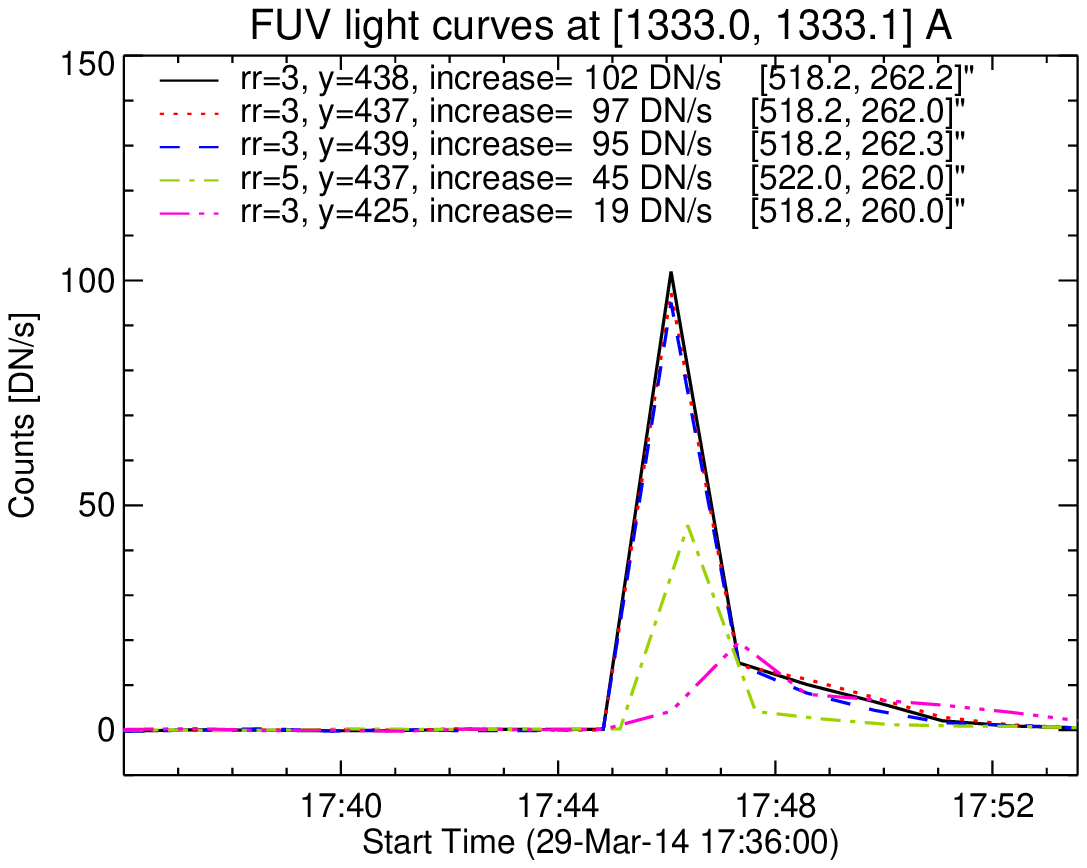}
   \includegraphics[width=.5\textwidth]{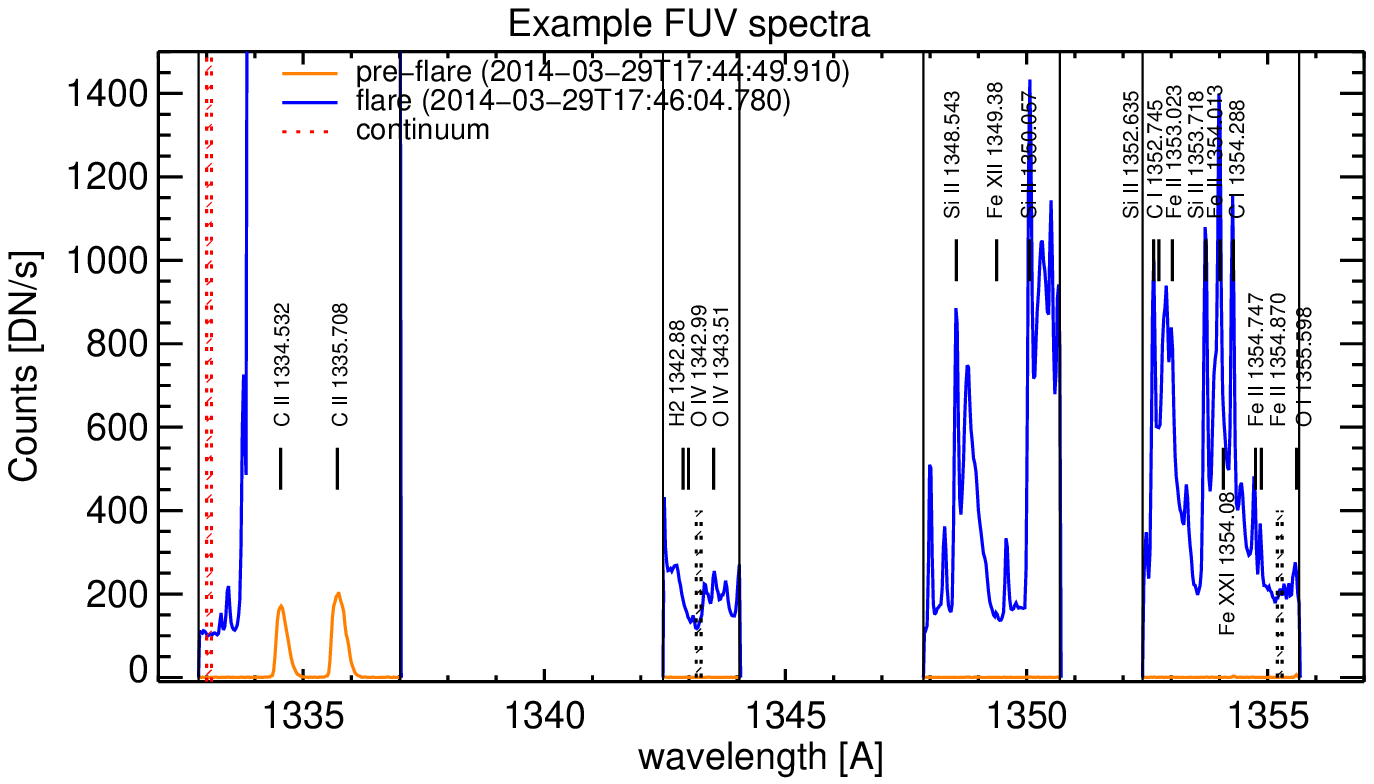}
	 \includegraphics[width=.47\textwidth]{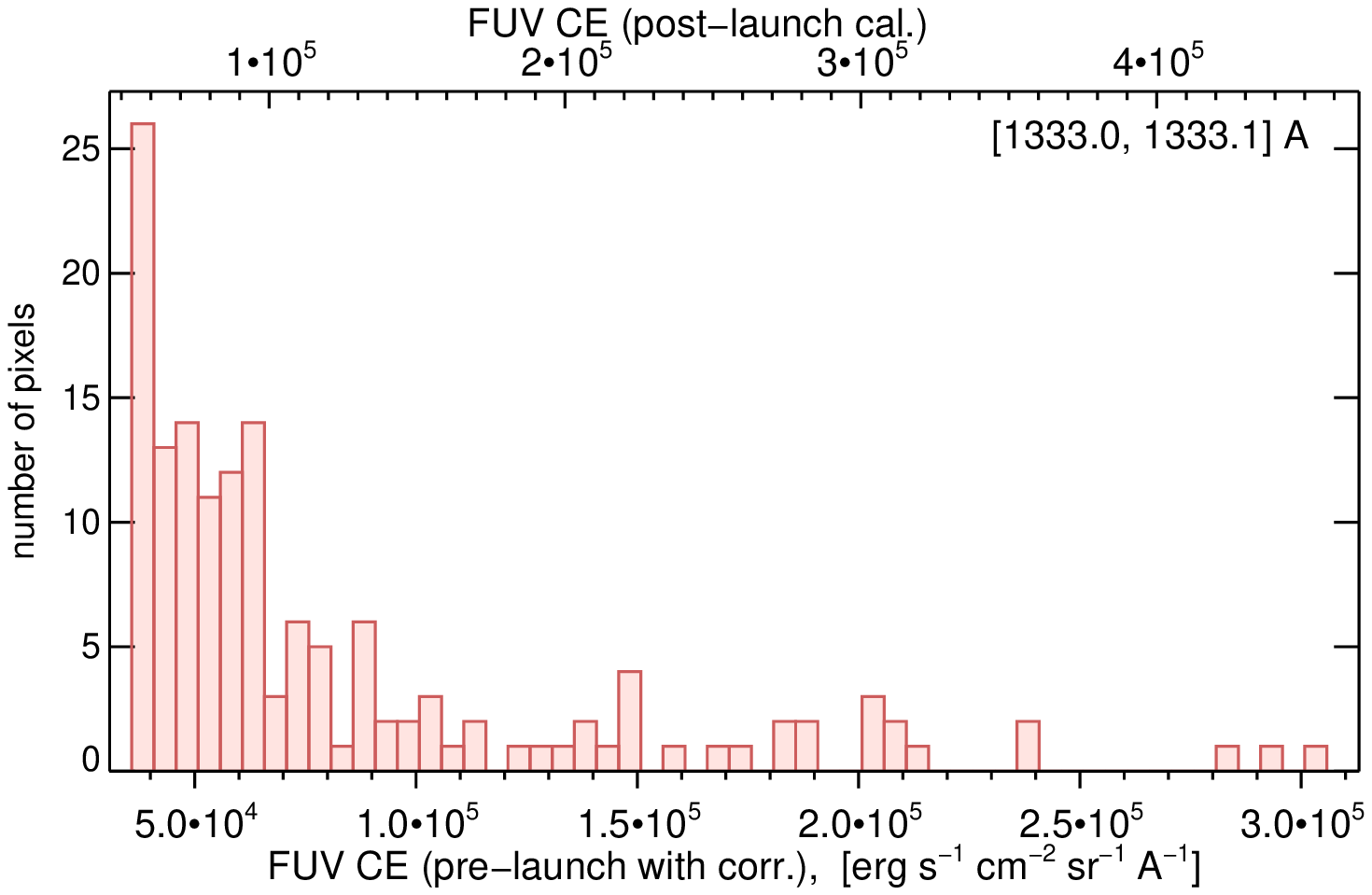}
   \caption{Top left: Locations of FUV CEs above 12 DN/s are marked in time-coded colors on the IRIS raster (slits indicated in blue).  The background is a negative image of an IRIS SJI 1400 at 17:47:36 UT. Note that the ribbon was moving, which is why the yellow marks, whose maximum CE occurred earlier than this image, do not seem to coincide with the ribbon in the image shown. Top right: Light curves from five selected pixels (the three maximum CE and two lower CE). Bottom row: Left: Example pre-flare and flare spectrum with the chosen continuum range indicated with red dotted vertical lines. The other tested continua (with more influence by spectral lines) are denoted with dark vertical dotted lines. The gaps in the spectrum show where the CCD was not read out or saved. Right: Absolute CE $>12$ DN/s as histogram for the two different radiometric calibration options.}
        \label{fuvce}
  \end{figure*}

\subsection{FIRS}

The Facility Infrared Spectrometer \citep[FIRS,][]{jaegglietal2010} is installed at the Dunn Solar Telescope of the National Solar Observatory in Sunspot, NM. It is a slit-scanning spectrograph with full polarimetry, optimized for observations in the IR. While a more complete description of our observations can be found in \citet{judgeetal2014}, we focused here on one raster, taken from 
17:40:12 to 18:01:39 UT on 2014 March 29 with the wavelength range 10813.8--10852.6 \AA. This range includes the three \ion{He}{1} line components (10829.091, 10830.250, 10830.340 \AA), the \ion{Si}{1} line at 10827.089 \AA, and several spectral ranges that can be assumed to be ``quasi-continua'' with no strong spectral lines. Each FIRS raster step took $\sim$12 s, and was composed of 10 polarization modulation cycles of 1.2 s with single exposure times of 125 ms. The 40 $\mu$m wide slit was oriented east-west, scanning from south to north. The spatial sampling and raster step size are both 0\farcs3. High-order adaptive optics were used when taking data, but no image reconstruction technique was used. This is acceptable for the IR, because seeing variations are less pronounced in the infrared than in shorter wavelengths, but any measured CE should be considered a lower limit.

\section{Analysis}\label{analysis}
 One of our main questions is how the continuum is formed. To investigate whether it is directly related to HXR emission, we look at the timing between RHESSI emission and IRIS CEs. We also investigate continuum spectral intensities, using fits of blackbody spectra and non-LTE model calculations.  In general, all derived CE values are lower limits owing to the fact that we cannot assure that any instrument has truly resolved bright flare footpoint kernels.

\begin{figure*}[bt] % fig 5
  \centering 
   \includegraphics[width=.77\textwidth,bb= 0 9 566 273,clip]{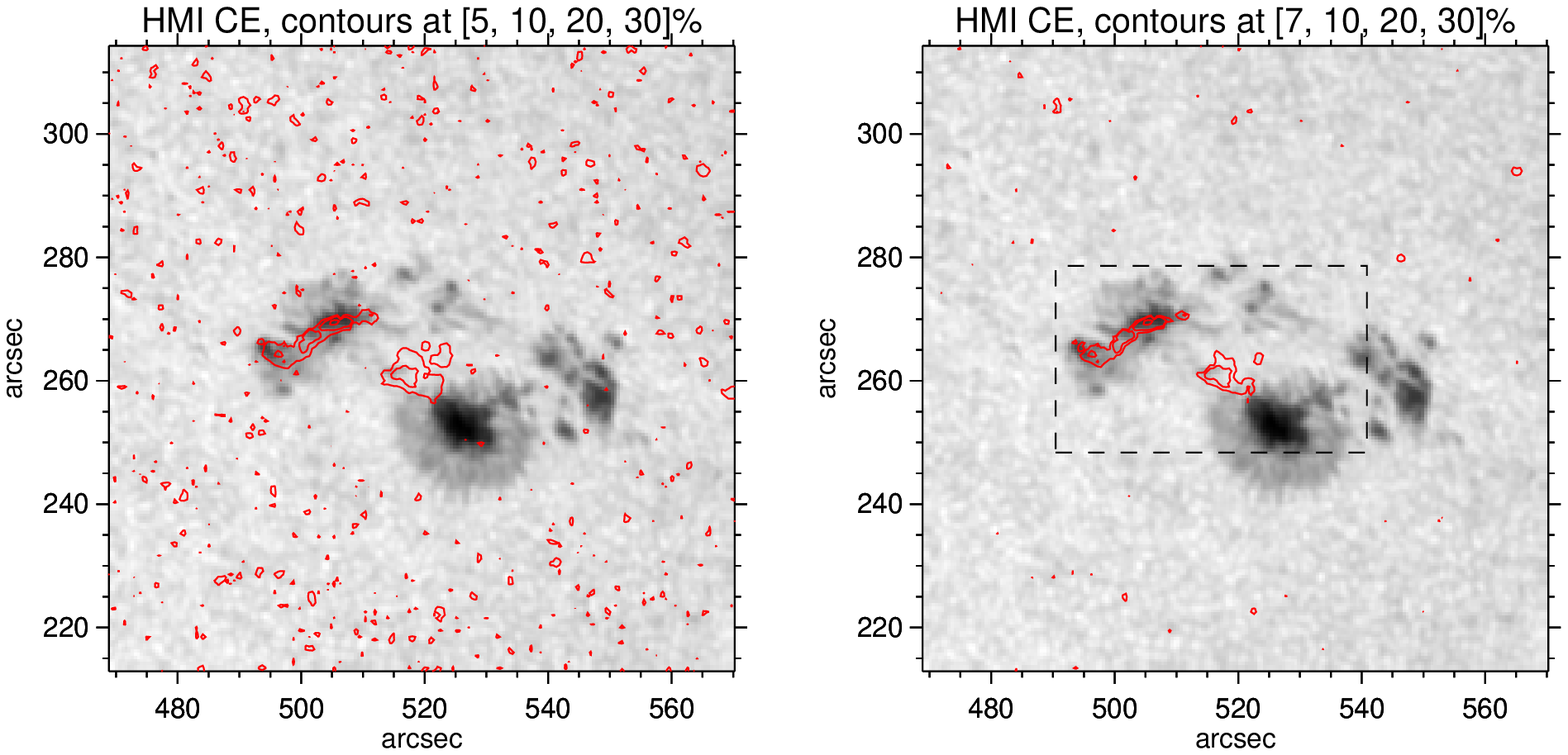}
   \vspace{-1mm}
   \includegraphics[width=.45\textwidth]{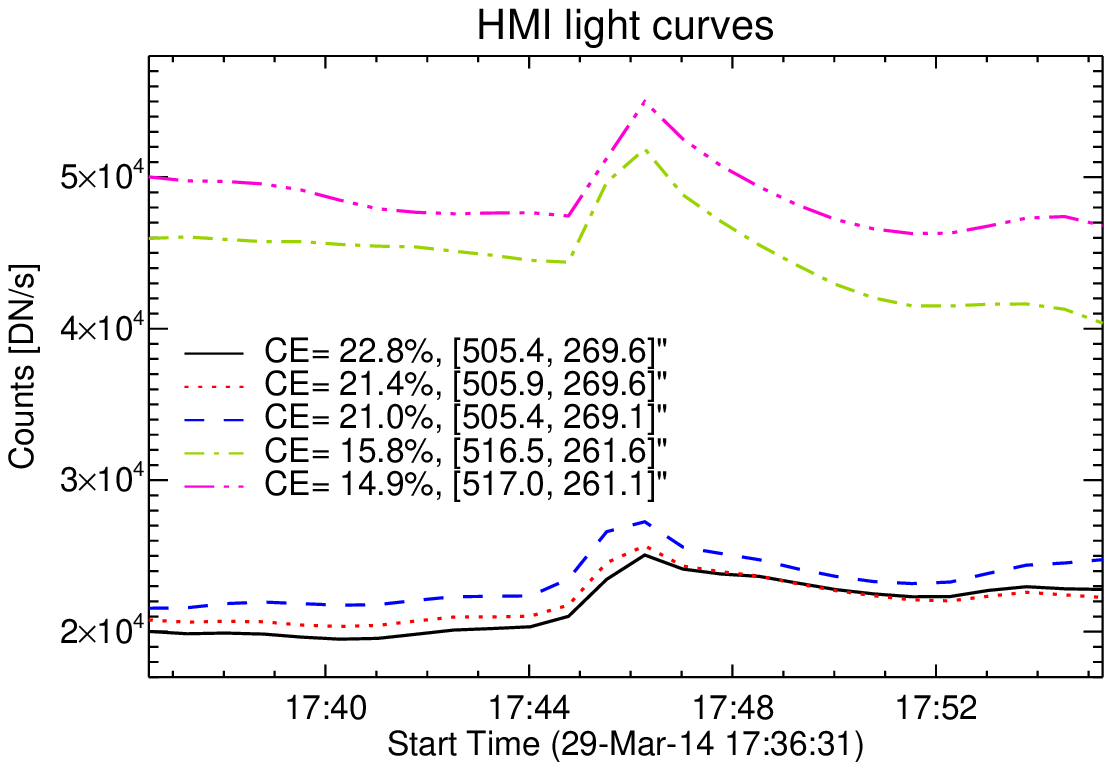}
   \includegraphics[width=.45\textwidth]{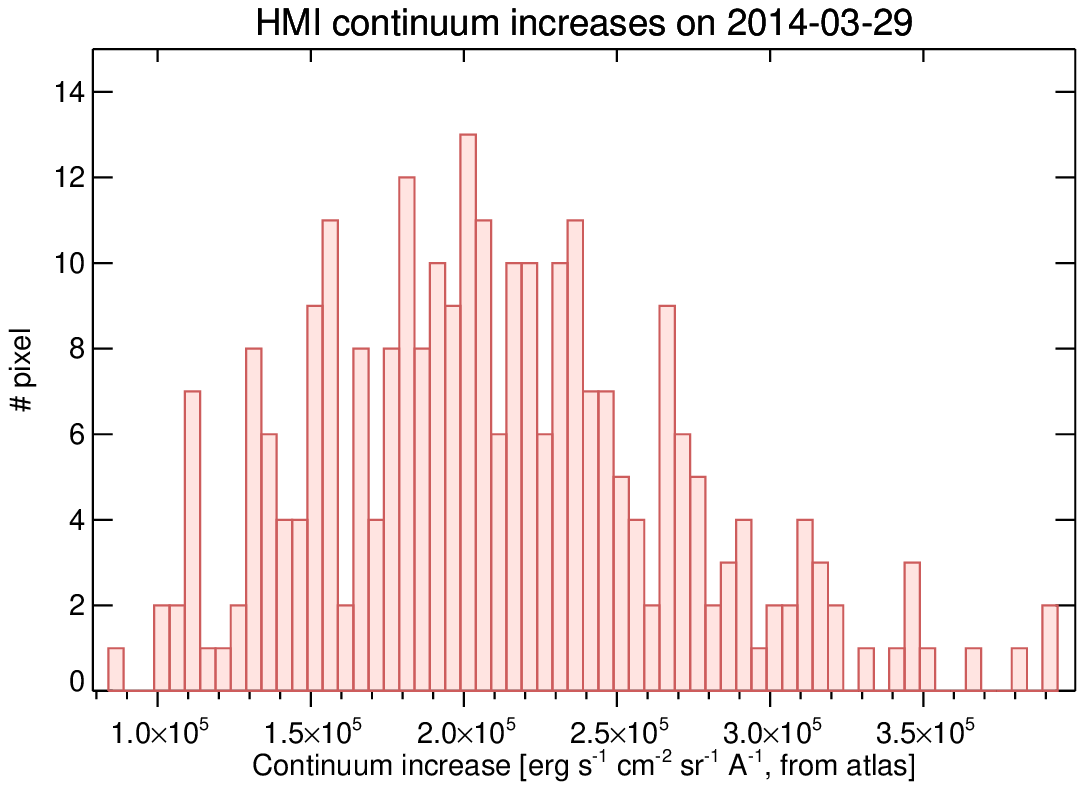}
   \caption{Top row: Contours of CEs on HMI continuum images showing why a cutoff of 7\% was chosen for flare-related intensity changes. The granulation may easily vary at 5\%. The dashed box denotes the selected area for the analysis. The HMI image and solar coordinates are from 17:46 UT.  Bottom left: Example light curves of the three maximum relative CEs, which occurred in the north-eastern footpoint and two examples from the south-western footpoint. Bottom right: Absolute values of the CE after a crosscalibration with the atlas from \citet{neckel1994atlas}.}
        \label{hmice}
  \end{figure*}

  \begin{figure*} % fig 1
  \centering 
   \includegraphics[width=.64\textwidth, bb=10 0 470 190]{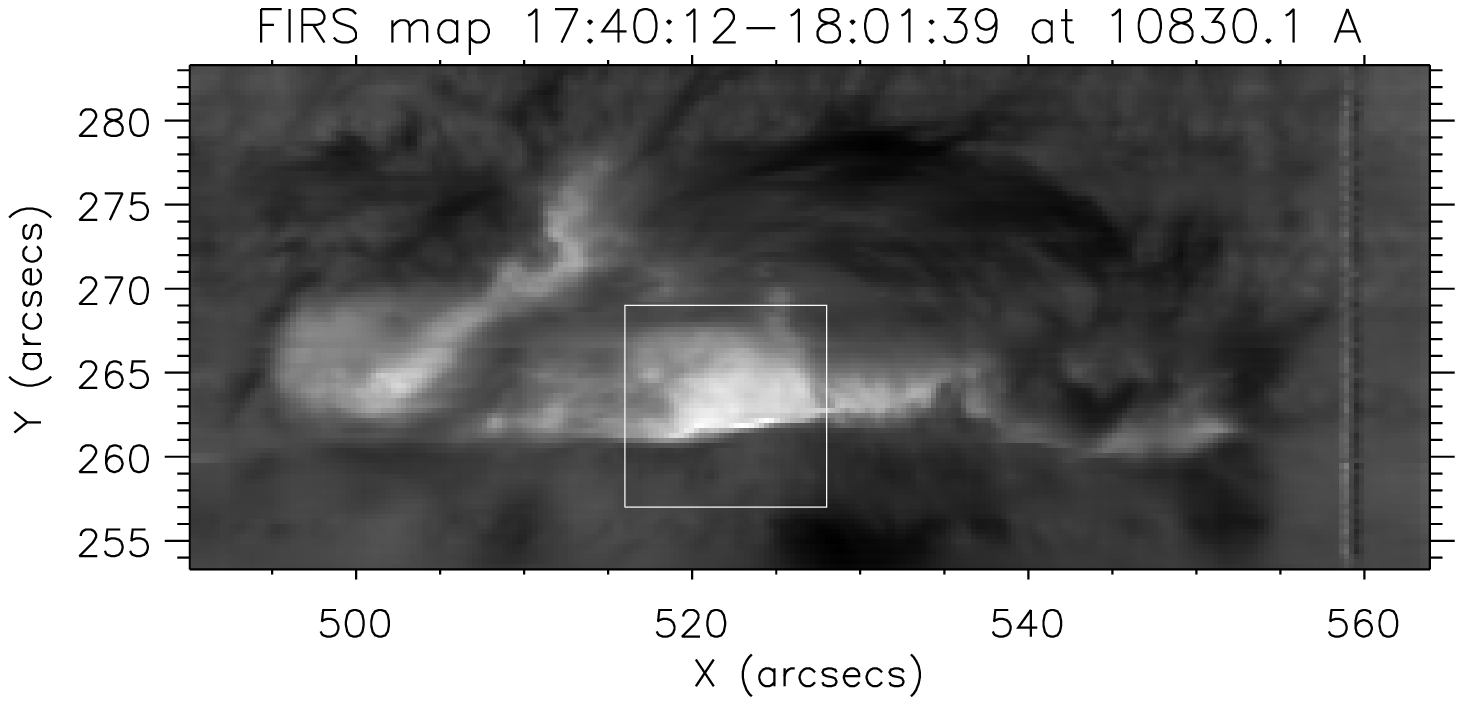}
   \vspace{-1mm}
   \includegraphics[width=.8\textwidth, bb=0 0 510 180]{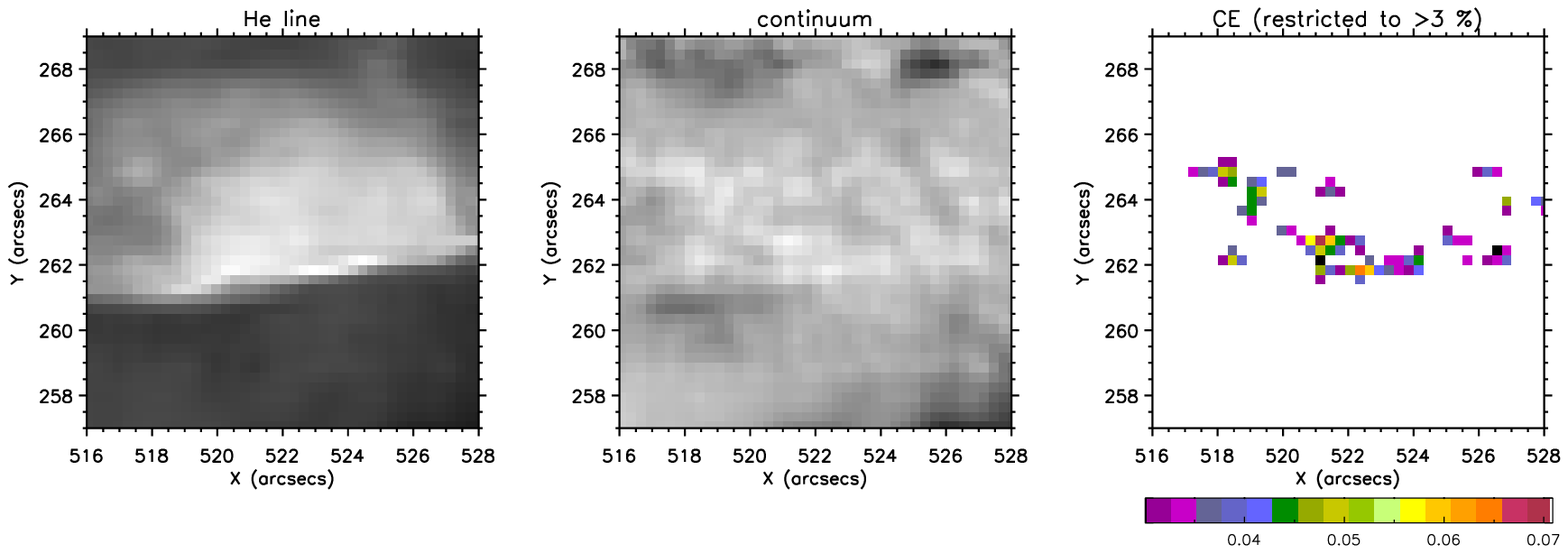}
   \caption{Top: FIRS raster at the wavelength of 10830.1 \AA\ (\ion{He}{1} line core). The raster was observed from S to N with the spectrograph slit oriented horizontally in the image. The box indicates the magnified area for the bottom row. From left to right: A magnification of the above raster, a continuum image, and the derived CE, relative to the quiet Sun, where they exceeded 3\%.}
        \label{firs}
  \end{figure*}
  
 \subsection{Statistics of NUV Continuum Enhancements}
  
IRIS NUV CEs during this flare were first identified by \citet{heinzelkleint2014}.
The line haze makes measurements extremely difficult, so instead we define a ``continuum'' window using only the range
2825.7--2825.8 \AA. Several small windows were tried and this one showed the lowest enhancements, suggesting the deepest formation layers and 
lowest influence of spectral lines. Then, we identify the raster $n$ when the continuum count maxima occurred ($I_{\rm flare}$, within a few minutes around 17:47 UT). From these maxima we subtract ``pre-flare'' continuum counts ($I_{\rm pre-flare}$), defined as the average of counts during rasters $n-4$ to $n-2$, to obtain the CE (in [DN/s]). This is done for each raster step separately, to avoid mixing spatial information, and therefore the temporal sampling of our reported CEs is 75 s, the duration of one raster. We selected a threshold of $I_{\rm flare}$ - $I_{\rm pre-flare} \ge$ 20 DN/s by verifying that most of the CEs occur in locations in the flare ribbons and that we do not include the generally smaller quiet Sun variations in our analysis.  Relative CEs were calculated as CE$_{\rm rel}$=(CE/preflare continuum),  plotted as a histogram in Fig.~\ref{ex} (bottom left).  The CE$_{\rm rel}$ values are usually below 200\%, with few outliers up to $\sim$430\%. Some example light curves including of the three highest outliers, are plotted in the top right panel. 
The variations indicate a steep (temporally unresolved) increase, followed by a slower decay. The absolute values for our measured CEs are shown in the histogram (bottom right), the x-axis labeled separately for the post-launch calibration and the pre-launch calibration with throughput correction. 

\subsection{IRIS FUV continuum enhancement}

In the FUV, the IRIS counts outside of spectral lines are usually close to zero, while they may increase to above 100 DN/s during a flare. 
It is therefore not possible to derive a relative enhancement from IRIS data.
%without drawing upon data from other instruments obtained 
%at different epochs {\color{red} LK for Phil: did these instruments show counts here? if yes, how (considering their worse resolution)?}. 
Using equation~(\ref{eq:cal}), and a typical quiet Sun 
absolute intensity of $\approx 15$ erg s$^{-1}$ cm$^{-2}$ sr$^{-1}$ \AA$^{-1}$ at 1330~\AA\ from 
\citet{vernazzaetal1981}, we find a count rate 
of 0.004 DN/s. %Phil: 0.016 DN/sec.  
Thus CE$_{\rm rel}$ for the IRIS FUV channel is a factor of $\le$25000.%6000$.   
We determine the absolute enhancement using the radiometric IRIS calibration. Locations where absolute 
FUV CE were found are shown in Fig.~\ref{fuvce} as colored bars overlaid on an IRIS SJI 1400 image with the 8 slit positions denoted with blue vertical dotted lines. Note that the background image is only a snapshot of a certain time (17:47:36 UT), while the different FUV CE occurred at different times, mostly when the ribbon was closer to them.

Another issue in FUV is to find ``continuum'' windows, because there are many (unidentified) spectral lines that appear during flares. Therefore, we selected several different windows where no obvious lines appeared during the flare and finally chose the one with the lowest values of CE (1333.0--1333.1 \AA, see Fig.~\ref{fuvce}). The pixels were found in a similar way as for NUV, with the requirement of having a CE of at least 12 DN/s. The bottom right shows CEs for the pre- and post-launch calibration. The derived FUV CE are about an order of magnitude lower than the NUV CE. 

%Both examples are from FUV1, because the spectral window on the other half of the CCD (FUV2, the longer wavelengths) suffered from severe overexposure and intensity bleeding from the \ion{Si}{4} line, rendering even ``continuum'' wavelengths unusable during the time where maximum CEs in FUV1 were observed. 

%\begin{figure*} % fig 4
 % \centering 
  % \includegraphics[width=.47\textwidth]{fuv_cont_lc_examples_2.eps}
  % \caption{Top row: Absolute CE of all pixels within two FUV ``continuum'' windows (given in titles). Only pixels with a CE of at least 20 DN/s were selected for the plots. Bottom row: .}
     %   \label{fuvce}
 % \end{figure*}

\subsection{HMI Continuum Enhancement}\label{sec:hmice}

We analyzed CEs of HMI continuum iñmages \textit{(hmi.Ic\_45s\_nrt)} obtained at 6173 \AA.  Solar rotation was removed 
and light curves were derived. We used the same method as for IRIS to derive $I_{\rm flare}$ and $I_{\rm pre-flare}$ and thus CEs. 
%For a maximum intensity occurring in image $n$, the pre-flare values were defined as the average of images $n-4$ to $n-2$. 
The top row in Figure~\ref{hmice} shows a map where HMI CEs were found. The left panel shows that intensity changes of the order of 5\% are common, even in granulation. We therefore selected a cutoff of 7\% for the flare-related changes (right panel), and also restricted the area that we analyze to the dashed box.
 
HMI does not have an absolute calibration, so we compared HMI disk center intensities with the atlas of  Brault \& Neckel \citep{neckel1994atlas}. At 6173 \AA, the atlas disk center intensity is
0.315$\times$10$^{7}$ erg s$^{-1}$ cm$^{-2}$ sr$^{-1}$ \AA$^{-1}$. Taking ten images around the flare maximum and averaging about 10\arcsec$\times$10\arcsec\ at disk center, we derive the HMI disk center intensity to be $\sim$60000 DN/s with a standard deviation of 300 DN/s. This gives us a simple conversion factor $I_{\rm abs}$ = $I_{\rm meas}$ $\times$ 0.315$\times$10$^{7}$ / 60000. Some example light curves of HMI are plotted in the bottom left panel of Fig.~\ref{hmice}, while  absolute CE are shown on the bottom right. 

HMI also has information about the north-eastern footpoint, which contains the strongest relative CE of up to 23 \% in HMI. The south-western footpoint has a maximum CE of 16\% in HMI. %The average absolute enhancements are about 2 $\times$ 10$^5$ erg\,s$^{-1}$\,cm$^{-2}$\,sr$^{-1}$\,\AA$^{-1}$, with a maximum up to 3.9 $\times$ 10$^5$ erg\,s$^{-1}$\,cm$^{-2}$\,sr$^{-1}$\,\AA$^{-1}$. 
We emphasize that these values are not directly comparable to IRIS data  (Figs.~\ref{ex} and \ref{fuvce}), because of the different spatial and temporal resolutions and instrumental profiles.

\subsection{FIRS Continuum Enhancement}\label{sec:firsce}
To derive CEs from FIRS data, because of variable seeing conditions, care has to be taken to compare stable raster steps. For each of the 100 raster steps, we derived an average continuum value of the quiet Sun away from the active region, and used it to normalize each raster step. This quiet Sun variation correlated well with broad band light level variations recorded simultaneously at the Dunn Solar Telescope, which could also have been used for a normalization. However, this only corrects for the variable throughput, not for the variable ``smearing'' due to seeing.
There is no pre-flare raster of exactly the same location under the same conditions, and therefore the ``pre-flare'' level was defined as the averaged intensity of the quiet Sun at several locations outside the AR, which appeared to have similar seeing as the flare ribbon location. Also, since the raster was built from S to N, and the HXR emission of that footpoint was moving N to S, the southern part of the FIRS raster was taken before the flare, there is an intersection point where HXR and FIRS were co-spatial and co-temporal, and the northern part of the FIRS raster was taken already in the decaying flare phase, making the interpretation of the observed CE more complicated. Additionally, there seems to be some vignetting or non-uniform transmission in the FIRS spectral direction, making the continuum level across the spectra tilted.

We examined two different spectral ranges to investigate the effect of the  ``tilted'' continuum. The first range was defined at 10823--10824 \AA. From Brault \& Neckel's atlas, we obtain a calibrated disk center intensity of 0.1035$\times$10$^7$ erg\,s$^{-1}$\,cm$^{-2}$\,sr$^{-1}$\,\AA$^{-1}$. From our flatfields, which were taken at disk center  right after the flare, we obtain average counts of 4500--5200 DN in that wavelength range. Our conversion factor is therefore $I_{abs}$ = $I_{\rm meas}$ $\times$ 0.1035$\times$10$^{7}$ / 4500 for an upper limit and $I_{\rm meas}$ $\times$ 0.1035$\times$10$^{7}$ / 5200 for a lower limit. Obviously, FIRS' spatial resolution and the seeing also limit the maximum detectable CE, so the upper limit should be interpreted with care. 
Similarly, for our second continuum window at 10840.1 -- 10840.4 \AA, we take the same atlas intensity and average flatfield counts of 5300--6000 DN.

Figure~\ref{firs} shows an image of the raster at 10830.1 \AA\ (He I line core) on the top with the white box denoting the cutouts in the bottom row. From left to right, these are: a magnification of the white box above, the continuum image in our first continuum window, and the relative CE with respect to the quiet Sun where only values above 3\% are shown to avoid the very noisy background. The maximum enhancement reaches 7\% in one pixel, which corresponds to CE=334 DN and CE$_{abs}$ = [6.13--7.09$\cdot$10$^4$, 6.45--7.30$\cdot$10$^4$] erg\,s$^{-1}$\,cm$^{-2}$\,sr$^{-1}$\,\AA$^{-1}$, for the two selected continuum ranges, respectively. The tilted continuum therefore does not seem to have much influence on the derived CE. Most pixels however showed much smaller enhancements, as can be seen in the figure.

 \begin{figure*} 
  \centering 
    \includegraphics[width=.95\textwidth,bb=0 0 566 197,clip]{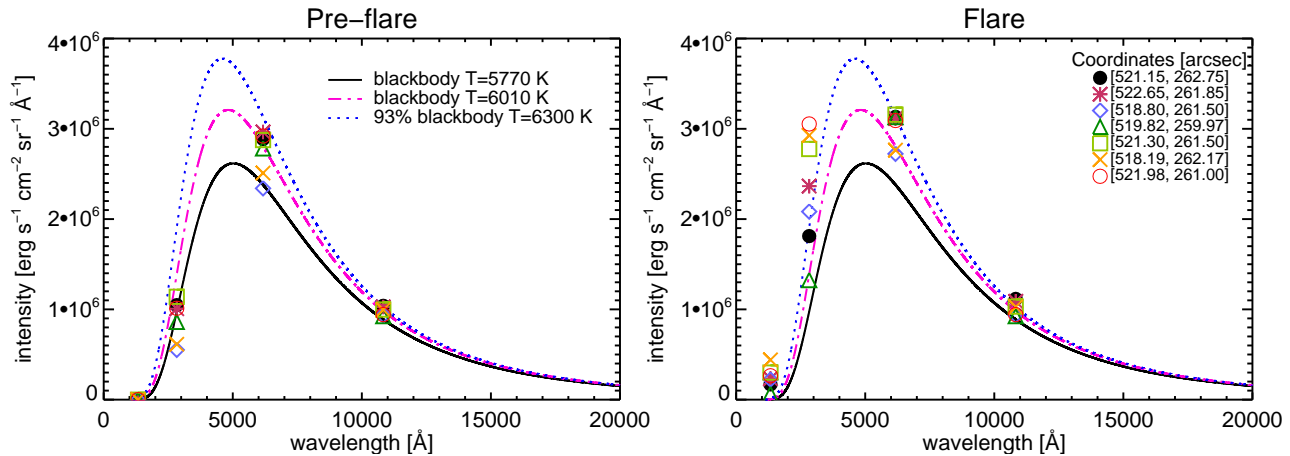}
 \caption{Left: Pre-flare intensities at different solar coordinates, which generally follow a blackbody of T$_{\rm eff} \approx$ 5770~K. Right: Intensities at the same solar coordinates during the flare. Especially in the NUV, they lie significantly above the blackbody spectra, indicating a different process (hydrogen recombination) for at least that part of the continuum.}
        \label{blackbody}
  \end{figure*}

%\begin{figure} 
 % \centering 
  %  \includegraphics[bb=400 0 720 360, width=.45\textwidth]{ce.eps}
 %\caption{{\color{red} suggest to keep only right panel. Remove G-band. Plot in Angstrom and same font as previous fig.}}
  %      \label{cephil}
 % \end{figure}

\section{Comparison of different wavelength bands: blackbody fitting}

The quiet Sun solar spectrum approximately corresponds to a BB spectrum with T $\approx$ 5770~K. This approximation becomes worse towards the UV. Our measured pre-flare intensities, seen on the left in Fig.~\ref{blackbody}, are a relatively good approximation of the BB, taking into account that solar NUV and FUV intensities are expected to lie below the BB spectrum. 

The figure shows the intensities of selected solar coordinates (patches of $\approx$0\farcs5$\times$0\farcs5) denoted by different symbols. This corresponds to one HMI pixel, the average of two FIRS pixels along the slit (when available at that location), and the average of 3 IRIS pixels along the slit (3$\times 0\farcs166$). Because of the raster step sizes, we only take one raster position, which, in case of IRIS, results in our ``average pixel'' of 0\farcs33$\times0\farcs5$. For NUV and FUV, we plotted the post-launch calibration values in the figure, because the intensities for the pre-launch calibration are higher. The intensities during the flare are shown in the right panel. 
For reference, some BB spectra for different temperatures are drawn, and they fit the visible and IR point quite well. The FUV flare data, formed in chromospheric plasma far higher than the other data shown owing to the high opacity of neutral silicon, differ dramatically from a single BB spectrum. In the NUV, the observed flare intensities are significantly higher than the BB. The NUV continuum probably arises from hydrogen recombination. Thus the Balmer continuum contributes a little to the opacity during the flare, leading to thick or thin emission below the continuum edge \citep{heinzelkleint2014}. To compete with the solar flux of radiation from the interior ($6\times10^{10}$ erg~cm$^{-2}$~s$^{-1}$), large amounts of radiation from the flare would have to emerge at unobserved wavelengths,
probably the Lyman continuum, Lyman lines, other UV and EUV lines of helium, or strong chromospheric lines of Ca~II, Mg~II or Fe~II.

%{\color{red} check intensities with new jumparr and write which cal was used or make error bars.}

\section{Modeling of Continuum}
  
To understand differences between BB spectra and the data, we employ
modeling. First, we test whether the existing flare atmospheric models
can explain our particular observations.  We focus on the grid of
theoretical flare models constructed by \citet{ricchiazzi1982} and
also use the semi-empirical FLA model of \citet{Mauasetal1990}.  

%As a second step, we compare the observations to 1D
%  non-LTE simulations with empirical model atmospheres using the RH
%  code \citep{uitenbroek2001,pereirauitenbroek2015}. Specific flare
%  codes like Flarix \citep{Varadyetal2010} and RADYN
%  \citep{allredetal2005} will be used as the next step in the future.

\begin{table*}[tbh]
\vspace{-4mm}
\caption{Balmer-continuum enhancement at 2830 \AA\  [erg s$^{-1}$ cm$^{-2}$ sr$^{-1}$ \AA$^{-1}$] for selected RC models}\label{tabpetr}
\begin{tabular}{l c c c  c | c c}
\hline
\centering
%a & a & a & a & a\\
      model     &     log(flux)    &   log(conductivity)   &   log(cor. pressure)  &  spectral index    &    Balmer enh.  \\
\hline\hline
%          E1        &      0         &      7           &          0           &          -            &             (quiet Sun reference)              \\
         E2        &      9          &     7            &         0            &         5            &          2.10 10$^4$\\
         E3          &   10         &      7           &          0           &          5           &           6.57 10$^4$\\
         E4         &    11         &      7           &          0           &          5           &           3.17 10$^5$\\
         E5         &     9          &     7            &         2            &         5            &          1.11 10$^5$\\
         E6         &    10         &      7           &          2           &          5           &           2.78 10$^5$\\
         E7          &   11         &      7           &          2           &          5           &           7.12 10$^5$\\
%         E8         &    10         &      6           &          0           &          5           &           8.82 10$^4$\\
%         E9         &    10         &      8           &          0           &          5           &           2.66 10$^2$\\
%         E10       &     10        &       7.5      &             2        &             5        &                 -\\
 %        E11       &     10        &       8         &            2         &            5         &             1.16 10$^3$\\
         E12       &     10        &       7         &            1         &            5         &             9.41 10$^4$\\
         E13       &     10        &       7         &            3         &            5         &             7.48 10$^5$\\
         E14       &     11        &       7         &            2         &            3         &             1.74 10$^6$\\
         E15       &     11        &       7         &            2         &            7         &             3.95 10$^5$\\
\end{tabular}
\end{table*}

  \begin{table*}
\vspace{-4mm}
\caption{Continuum enhancements in FLA model}\label{tab1}
\begin{tabular}{c c c c  c | c c}
\hline
\centering
%a & a & a & a & a\\
                  & quiet Sun (C7)       & FLA                & model abs. CE & model rel. CE & obs. abs. CE  & obs. rel. CE\\
                 &  \multicolumn{2}{l}{ [erg s$^{-1}$ cm$^{-2}$ sr$^{-1}$ \AA$^{-1}$]} \\
\hline\hline
HMI         &     3.24$\cdot$10$^6$  & 3.51$\cdot$10$^6$ & $\approx$2.7$\cdot$10$^5$ &  8\%    & $\approx$1-4$\cdot$10$^5$ & 7--23 \% \\
FIRS       &   1.06$\cdot$10$^6$ & 1.12$\cdot$10$^6$ & $\approx$0.6$\cdot$10$^5$ & 6\%  & $<0.7 \cdot 10^5$ & $<$7\% \\
\end{tabular}
\end{table*}

\subsection{Hydrogen Balmer and Paschen Continua}

We used a grid of 1D static flare models of \citet{ricchiazzicanfield1983} (hereafter referred to
as RC models), where the temperature 
structure is computed from the energy-balance between electron-beam heating, conduction and net 
radiation losses. Moreover, the models also account for an enhanced coronal pressure which is due
to the evaporation. This theoretical grid of ab-initio models depends on a few key parameters
which are the electron-beam energy flux at cut-off energy 20 keV (10$^9$--10$^{11}$ erg\,s$^{-1}$\,cm$^{-2}$), 
spectral index varying from 3--7, thermal conductivity parameter and the coronal pressure at
the upper boundary varying between 1--1000 dyn cm$^{-2}$. We used the non-LTE code MALI based on \citet{rybickihummer1991} and \citet{heinzel1995} to synthesize the
hydrogen recombination continua for all models of the grid and evaluate them at $\mu$=0.82. We included the hydrogen non-thermal collisional rates caused by the
electron beam, consistently with the RC models. 
While the hydrogen Lyman continuum is optically thick and its spectrum can be compared with
e.g. SDO/EVE observations, all subordinate recombination continua are thin within the chromospheric
layers where they are formed in RC models. This has the advantage that we can compute only the
chromospheric (i.e. flare) contribution and compare it directly with observed one taken as the flare
spectrum minus the pre-flare (or quiet-Sun) background. This was also done in a preliminary analysis of \citet{heinzelkleint2014}. The models and the computed CE at 2830\,\AA\ in radiometric units as the IRIS observations are listed in Table~\ref{tabpetr}. Our observed enhancement of  $\sim$0.5-2.0$\times$10$^6$  erg s$^{-1}$\,cm$^{-2}$\,sr$^{-1}$\,\AA$^{-1}$ is consistent with some selected RC models, namely E7, E13, and E14 (the numbering corresponds to the PhD thesis of P. Ricchiazzi, different from Table 1 of RC). Model E14 with a beam flux 10$^{11}$ erg  s$^{-1}$ cm$^{-2}$ and the spectral index equal to 3 provides the largest enhancement
in the whole RC model grid and agrees with the flux derived from RHESSI observations (see Sect.~\ref{rhsec}), while the spectral index for this flare varied between 3.5 and $\sim$5 during the relevant times. Inspection of the table shows that these models require downward energy fluxes in excess of $10^{11}$
erg\,s$^{-1}$\,cm$^{-2}$, or in excess of $10^{10}$ erg\,s$^{-1}$\,cm$^{-2}$ under conditions of high coronal pressures. However, the other recombination continua (Paschen, Brackett) are too weak in these RC models to explain the enhancement in HMI and FIRS we do observe.

  \begin{figure*} % fig 1
  \centering 
   \includegraphics[width=.87\textwidth,bb=5 5 566 420,clip]{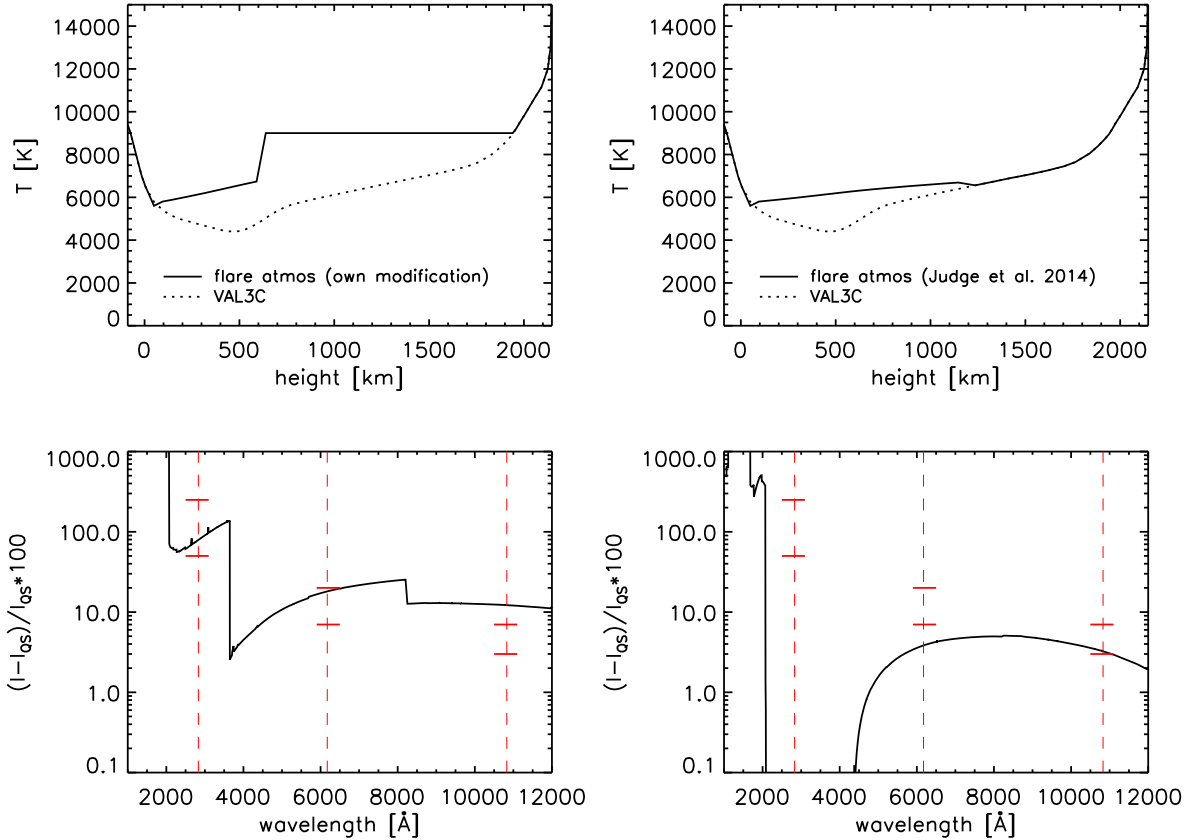}
   \caption{Top row: Temperature structure of two semi-empirical model atmospheres (solid) and of the quiet Sun (VAL3C, dotted). Bottom row: Resulting enhancements of the spectra relative to the quiet Sun from UV to IR, plotted in percent on a log-scale. The model on the left fits the observed CE, whose ranges are indicated by red bars, relatively well.}
        \label{philatmos}
  \end{figure*}

\subsection{HMI and FIRS Continua}

To account for the observed enhancement of the continua in the visible and IR, we considered the semi-empirical 
white-light flare model FLA of \citet{Mauasetal1990}. Based on this model, \citet{HeinzelAvrett2012}
computed the synthetic spectrum of the visible to radio continua. The intensities at HMI (6173 \AA) and FIRS
(mean wavelength between the two continuum windows from Sec.~\ref{sec:firsce}: 10832 \AA)
wavelengths are in the range of our HMI and FIRS observations, are shown in Table~\ref{tab1} and were calculated for $\mu$=1.

As a reference, we used the semiempirical quiet-Sun model C7 of Avrett and
Loeser (2008). According to \citet{HeinzelAvrett2012}, this is possible because C7 and FLA were constructed independently and FLA has an enhanced photospheric temperature which causes, mainly due to H$^-$,
the enhanced white-light continuum. C7 is the mean quiet-Sun model based on SOHO/SUMER UV-spectra. The continuum intensities from C7 are slightly higher compared to the Brault \& Neckel atlas values in Sect.~\ref{sec:hmice} and ~\ref{sec:firsce} and using the atlas intensities would increase the calculated relative CE in Table~\ref{tab1} by $<3\%$.

Any 'hotter' model, e.g. representing facular regions, will lead to lower contrast with respect to FLA. The FLA model matches the observed CE for HMI and FIRS well, both in absolute and relative CE. The larger observed relative HMI CE values (up to 23\%) may result from the fact that the CE was observed in and near sunspots, where the quiet-Sun model probably does not represent the pre-flare state very well.

\subsection{FUV continuum from IRIS}

It is known that this continuum and other UV continua may substantially increase during flares, seen both in observations \citep[e.g.,][]{brekkeetal1996} and models \citep[e.g. flare models F1 to F3 in][]{avrettetal1986}.
As already discussed by \citet{heinzelkleint2014}, this is the UV continuum formed above the
temperature minimum (lower chromosphere), due to Si I. 
These continua have no direct
relation to the previous ones, but the heating in these chromospheric layers can be due to the Si I opacity \citep{Machadoetal1986}. We postpone the detailed modeling of the IRIS/FUV continuum to another paper.

 \subsection{Non-LTE modeling with RH}
 
% {\color{red} Issue: QS VAL3C atmosphere fits the flare data. What should we conclude here? Do simulations with increased opacity have lower intensity?}
 
 To investigate the influence of an empirically derived temperature structure on the resulting continuum, we
 modeled the spectrum from UV to IR using the RH code \citep{uitenbroek2001,pereirauitenbroek2015}, which solves the non-LTE statistical equilibrium equations for selected atoms. Here we select H, C, Si and Fe, identically to \citet{judgeetal2014}, and apply a similar procedure of empirically varying an input atmosphere by introducing temperature variations (photospheric and chromospheric temperatures are enhanced). While the results are not unique and different temperature models may lead to similar spectra, we can nevertheless investigate the dependence of the continuum on the height where the temperature structure is changed. These calculations are done for $\mu=$0.77, close to the observed $\mu=$0.82. %{\color{red} Petr, yes, this can be changed, but it's a pain (crashed in my case).}

  \begin{figure*} % fig 1
  \centering 
   \includegraphics[width=.9\textwidth]{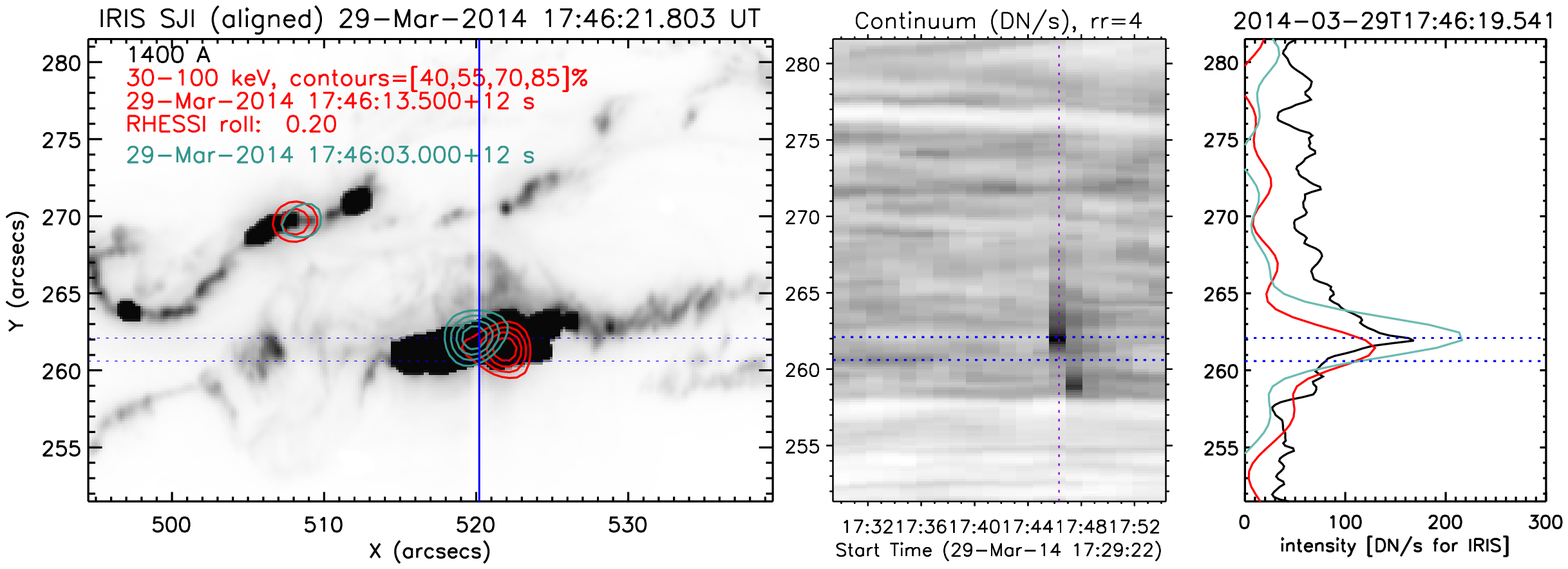}
   \includegraphics[width=.9\textwidth]{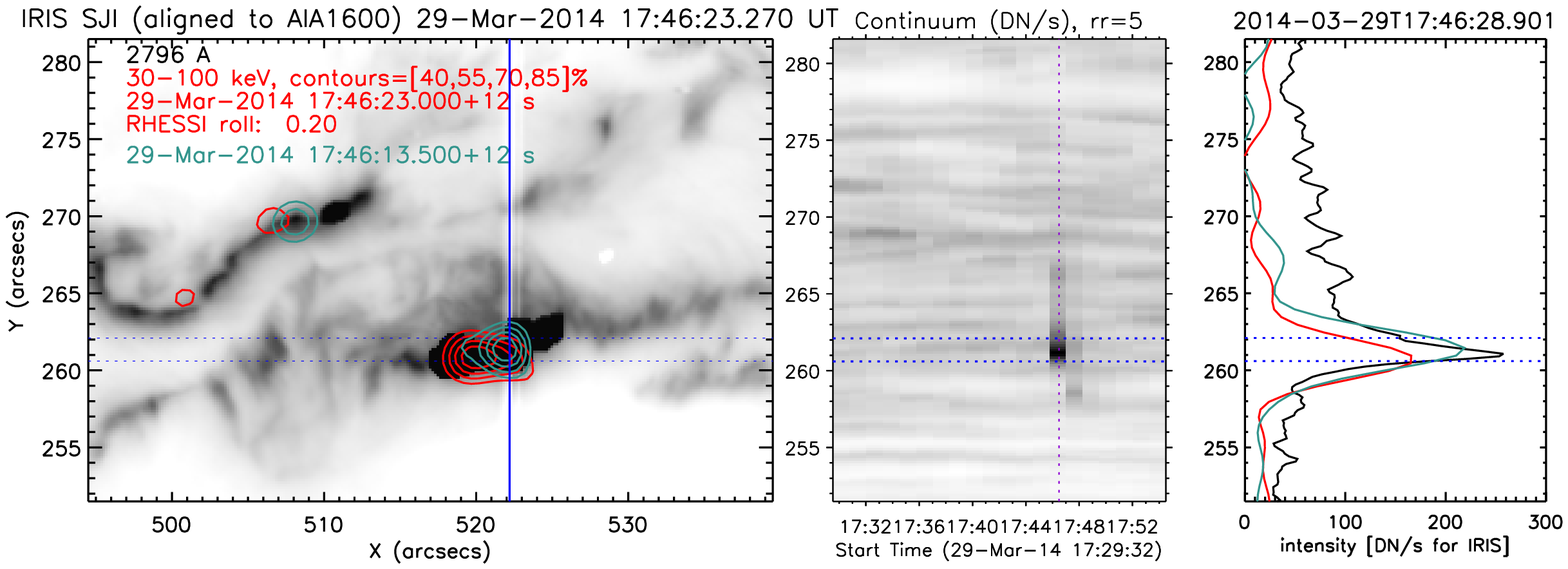}
   \caption{Left column: RHESSI contours (red; light blue for 10 s earlier) on color-reversed IRIS SJI with the slit marked blue. Middle column: Color-reversed continuum intensity evolution showing the continuum emission in black. The dotted vertical purple line indicates the location where the intensity for the plots in the right column was derived. Right column: Intensity cross-section at the given raster step and time (see image titles). The red line (light blue line for 10 s earlier) is the RHESSI intensity cut (in arbitrary units) at the same time and solar $X$. The maxima of the IRIS continuum and RHESSI HXR emission nearly coincide.
   The horizontal dashed lines are for reference to indicate the same locations in all three panels. 
   %The middle row shows the effect of a different RHESSI roll angle correction, with 0.1 degrees instead of 0.2 degrees.
   }
        \label{rhiris}
  \end{figure*}

The calculated quiet Sun spectra generally have too high intensities at all wavelengths, especially in the UV, which indicates that we have not included sufficient opacities (as there are no background opacities). We circumvent this problem by comparing relative CE values. Fig.~\ref{philatmos} shows two selected resulting atmospheres. The top row shows their temperature structure (solid lines), with the quiet Sun for comparison (dotted line). The bottom row shows the relative enhancement with respect to the quiet Sun. The red dashed lines indicate the observed wavelengths and the red horizontal bars show the range of observed CE. Clearly, the left model reproduces the observations relatively well and the lower observed enhancement in IR may be explained by the seeing and the relatively slow scanning of the spectrograph, which may smear stronger CE. Strong hydrogen recombination jumps are visible, which arise from the enhanced chromospheric temperatures in this model. We found that both, a small increase at photospheric temperatures and a strong increase at chromospheric temperatures is required for a good fit. The model on the right, which was the best fit to the observed \ion{Si}{1} 10827 \AA\ line in \citet{judgeetal2014}, has a lower temperature enhancement compared to the model on the left, which explains the absence of any hydrogen recombination jumps in the lower right panel. While both atmospheres fit the wing of the Si line, our new model (left column) produces a very high emission of the Si line core (factor $\sim$1.5 higher), which is higher than any of the observations. Another possible problem of the left model is the presence of the Paschen jump around 8200 \AA\ of $\sim$ 10 \% enhancement, which should be observable, but to our knowledge, the Paschen jump has never been observed. 

\textit{The conclusion from all these modeling efforts is that temperature changes \textbf{both} in the photosphere and in the chromosphere are required for the observed magnitude of CE. }

%{\color{red} PLOT CONTRIBUTION FUNCTION!}

%{\color{red} todo: lines in fig to go to 0, make lines gray, continuum black}

 \section{Timing and Energies of HXR and Continuum Emission}
 
 Apart from the spectral behavior of the continuum analyzed above, we can also investigate its temporal behavior and its relation in space and time to the accelerated electrons detected by RHESSI. This allows us to measure the delay in the production of the continuum emission and compare the energy input from accelerated electrons to the emitted radiation.
 
\begin{figure*}[htb] % fig 1
  \centering 
   \includegraphics[width=.85\textwidth]{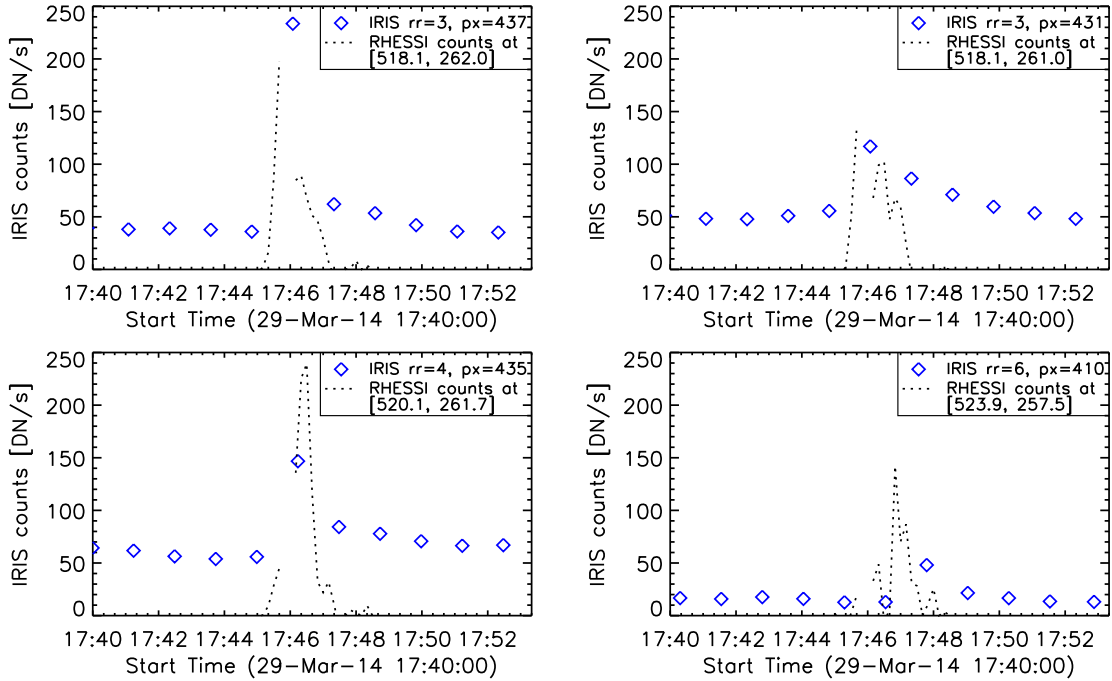}
   \caption{Light curves at different solar coordinates for IRIS (with a cadence of 75 s due to the raster) and RHESSI (10 s cadence with 12 s integration time). The RHESSI units are arbitrary, but consistent between the different plots. Missing RHESSI values around 17:46 UT are because of an attenuator state change.}
        \label{lc}
  \end{figure*}

 \subsection{Temporal correlation between NUV continuum and HXR}
 
 We overplotted the RHESSI contours on IRIS data and found two times, where the IRIS slit crossed the HXR source. These images are shown in the left column of Fig.~\ref{rhiris}. The \textit{date\_obs} (start time of observation) of the SJI is plotted in the title of these images. The intensity of the SJI is reversed, simultaneous RHESSI contours are shown in red, 10 s earlier contours in light blue and the IRIS slit is depicted in dark blue.
 %The middle row shows the effect if the RHESSI roll correction were only 0.1 degrees. Clearly, the eastern footpoint then no longer coincides with the ribbon, indicating that 0.2 degrees are a more logical choice. %The bottom row may benefit from a roll correction of 0.25 degrees to perfectly match the eastern ribbon, but this is within the error bars because they are only contours of 40\% and for consistency, we keep 0.2 degrees.
  
 The middle column shows the temporal evolution of the NUV continuum intensity (colors reversed). A clear emission signal appears at $\sim$17:46:20 UT, and a weaker signal is visible slightly further south at $\sim$17:47:30 UT. The index of the raster step is shown in the title (e.g. rr=4 is the 5th of 8 steps). The vertical dotted line indicates the time when the intensity for the right panels was plotted. The right column shows the NUV continuum intensity along the slit in black. The time in the title corresponds to the start time of the NUV exposure, which lasted 2.4 s. The red line is the arbitrarily scaled RHESSI emission at this time and solar $X$ coordinate and the light blue line the RHESSI emission at an earlier time (given in left panels). The maxima of the co-temporal RHESSI intensity are slightly further south than those of the continuum intensity, but the difference is less than one arcsecond. While such a behavior would be expected, as the accelerated particles are probably hitting the chromosphere first, and the continuum emission appears subsequently, our measured difference between the two curves is within the 
 %RHESSI resolution limit (2.3\arcsec\ in the best case) and within our 
 alignment errors ($< 2\arcsec$). We can however conclude that the timing between RHESSI centroid and subsequent maximum continuum emission must be less than 15 s. This is the timespan at which the RHESSI HXR footpoint emission, which was moving, coincided with the maximum continuum emission location.

Another way to investigate the timing between HXR and continuum emission is to create light curves of both at selected solar coordinates. Some examples are shown in Fig.~\ref{lc}. The diamonds indicate the IRIS CE, and the dashed lines are RHESSI counts (arbitrary units, but same scaling for all plots). The RHESSI images used for this plot were created with intervals of 10 s and integration times of 12 s. There was a RHESSI attenuator state change at 17:46, making count rates around that time unreliable and they were omitted from the plots. The IRIS raster cadence is not fast enough for an analysis of the order of less than one minute, but in general, the peak times of CE and RHESSI agree well. It is also visible that generally, the stronger the RHESSI emission, the stronger the continuum seems to increase. For the cases where this is not true, it is possible that higher CE was present between two IRIS raster times. The decay of the enhanced continuum emission is slower and persists after the HXR signal disappears. This could be related to the relaxation times for hydrogen recombination.

\begin{figure}[htb] % fig 1
  \centering 
   \includegraphics[width=.5\textwidth]{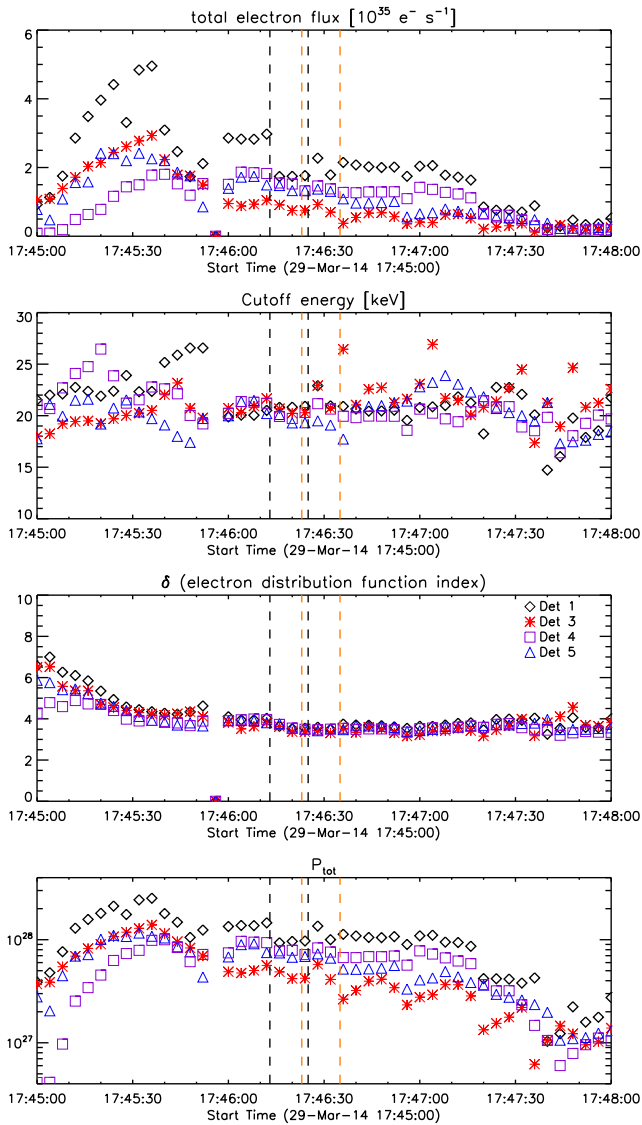}
   \caption{Parameters derived from fits to RHESSI energy spectra during the impulsive phase of the flare. Plotted are (from top to bottom) the total electron flux, the fitted cutoff energy, the spectral index of the non-thermal electron distribution and the calculated total non-thermal power. Vertical lines indicate the two time ranges that were used to reconstruct the RHESSI contours in Fig.~\ref{rhiris}}
        \label{rhfits}
  \end{figure}

\subsection{Comparison of Input and Output Energies}\label{rhsec}
 \subsubsection{Energy derived from RHESSI}
To estimate the energy deposition rate by flare-accelerated electrons within the area of the IRIS slit, we use standard RHESSI spectroscopy techniques \citep[e.g.,][]{linetal2002} to first estimate the energy deposition rate in the south-western footpoint. We fit the energy spectra during the impulsive phase with a thermal component, a thick-target model (power law) and a correction for pile-up. The result of these fits is shown in Fig.~\ref{rhfits} for different RHESSI detectors. Vertical lines indicate the two time ranges that were used to reconstruct the RHESSI contours in Fig.~\ref{rhiris}. The total power of the non-thermal electrons ($P_{\rm tot}$) was calculated as \begin{equation}
 P_{\rm tot} = \frac{\delta-1}{\delta-2} F_{\rm tot} \, E_{\rm c} \, {\rm [erg \, s^{-1}]}
\end{equation}
with the cutoff energy $E_c$ (1 keV = $1.6 \cdot 10^{-9}$ erg), the total electron flux $F_{tot}$, which depends on E$_c$, and the electron spectral index $\delta$. For the two time intervals where the slit coincided with HXR emission (see Fig.~\ref{rhiris}), the energy deposition rate of flare-accelerated electrons above a cutoff of $\sim$20 keV in the entire footpoint becomes $\sim 8\times 10^{27}$ erg s$^{-1}$. As the low energy end of the power-law distribution is hidden by the much stronger thermal emission at lower energies, these values are lower limits. To estimate the energy deposition rate in the area given by the IRIS slit with a slit width of 0.33\arcsec, the deposition within the entire footpoint has to be multiplied by the fraction of the area covered by the slit relative to the entire footpoint area. From deconvolved CLEAN images the estimated extent of the HXR footpoint normal to the slit is about $\sim4.4\arcsec$. Hence, the lower limit for the energy deposition by non-thermal electrons above $\sim$20 keV in the IRIS slit area becomes 5.5$\times 10^{26}$ erg s$^{-1}$. The dependence of these values on the low energy cutoff is
 \begin{equation}\label{eq3}
 \frac{P_{\rm tot}(E) }{P_{\rm tot}(E_c)} = \frac{E^{-\delta+2}}{E_c^{-\delta+2}} = \left( \frac{E}{E_c} \right)^{-\delta+2}.
 \end{equation}

A cutoff of $n$ keV ($n$=\{10, 40\}) would therefore increase/decrease the total power of the non-thermal electrons to ($n$/20)$^{-2} \approx \{400\%, 25\%\}$, respectively, of the above value.

To compare with the energy losses derived from the measured data, we convert into the same units by dividing by the area of the footpoint. We assume the width of the ribbon to be 1\arcsec, as it is unresolved in RHESSI. We calculate $8 \times 10^{27} / (4.4 \times 1 \times (7.25 \times 10^7)^2)$ = $3.5 \times 10^{11}$ erg s$^{-1}$ cm$^{-2}$. This calculation assumes that the energy is distributed equally everywhere in the ribbon. While this is probably not true, the IRIS slit intersects the RHESSI emission at the 50\% contour, and not at the maximum, which is why the calculation seems reasonable for our case.  In the next section, these rates are compared to the observed radiative losses to estimate the fraction of deposited energy that is lost by radiation in the continuum. It needs to be kept in mind though that an error in the (unresolved) footpoint area, then directly translates into the same error for the deposited energy.

\subsubsection{Energy Losses in the Continuum}

We calculate the difference of two integrals of blackbody spectra (``flare'' at 93\% of 6300 K and pre-flare at 5770~K, see Fig.~\ref{blackbody}) to estimate the flare energy radiated in the visible continuum. Knowing that the whole continuum emission cannot be described by a pure blackbody, we estimate the contribution from the Balmer continuum as a second step. 

By multiplying the difference of the two BB spectra by $2\pi$, thus assuming optically thick radiation, we can convert [erg s$^{-1}$ cm$^{-2}$ sr$^{-1}$ \AA$^{-1}$] to [erg s$^{-1}$ cm$^{-2}$], which results in an energy loss of $4 \times 10^{10}$ erg s$^{-1}$ cm$^{-2}$. This is only a simple approximation, because the radiative losses are not directly related to the emergent intensity in the optically thick case. A crosscheck with the losses from the FLA atmospheric model \citep[Table 3,][]{Mauasetal1990}, integrating them from the photosphere (0 km) to the temperature minimum region ($\sim$400 km), yielded a consistent result of $2.4 \times 10^{10}$ erg s$^{-1}$ cm$^{-2}$.

For the Balmer continuum (NUV), we calculated the losses from the E14 model and multiplied them by $4\pi$, assuming optically thin emission, giving $3.8 \times 10^{10}$ erg s$^{-1}$ cm$^{-2}$, which means that the losses in the Balmer continuum are comparable to those in the H$^{-}$ continuum.

Calculating the total loss depends on the assumed scenario. If we simply add the losses of both continua, the resulting energy loss is $\sim8 \times 10^{10}$ erg s$^{-1}$ cm$^{-2}$, which is 23\% of $3.5 \times 10^{11}$ erg s$^{-1}$ cm$^{-2}$.
This result can be understood as that the energy of 23\% of the electrons above 20 keV would be needed to produce the energy radiated in the continuum. Another possibility (using Eq.~\ref{eq3}) would be to require all of the energy of the non-thermal electrons above 40 keV for the observed CE. Because of the unresolved footpoint area, the error bars during the fitting of RHESSI spectra, and the unknown cutoff, the error bar of this result may be significant and cannot be reliably estimated. We are therefore cautious to overinterpret our result, but it is safe to say that the continuum radiation is not negligible in the overall energy balance of the flare.

If we assume that the electron beam is stopped in the chromosphere and deposits its energy there, only the Balmer continuum loss has to be compared with the energy deposit. I.e.
3.8 $\times 10^{10}$/ 3.5 $\times 10^{11}$, which is about 11 \% of deposited energy by electrons above 20 keV. In the backwarming scenario, we can assume that half of the Balmer emission is radiated away, while the other half ($\sim2 \times 10^{10}$ erg s$^{-1}$ cm$^{-2}$) shines down and may heat the photosphere. This energy could then be re-radiated by the photosphere, and its values are compatible to the losses calculated from FLA or the BB approximation, which supports the backwarming hypothesis, but does not allow us to conclusively determine the scenario.

%The cutoff energy of the non-thermal emission is unknown because it is hidden by the thermal component. The area of the footpoints and energy deposition is unresolved and under- or overestimating it by a factor of 10 will lead to the same factor of uncertainty in the result. It is also unknown how homogeneous the area is where the energy is deposited and considering the bright kernels at sub-arcsec resolution seen with IRIS, it probably is inhomogeneous. If the energy is concentrated at some locations, this may also lead to an error for the assumed energy within the IRIS slit. 

\section{Discussion and Conclusions}

In summary, we find the following
\begin{enumerate}
\item The observed flare continuum intensities in UV, VIS and IR during the X1 flare SOL20140329T17:48 do not fit a simple blackbody spectrum, with especially the UV being too high. This indicates that other processes, such as hydrogen recombination (e.g. Balmer continuum) contribute to the continuum emission.
%The average observed CEs in UV, VIS and IR during the X1 flare SOL20140329T17:48 can be described with a blackbody curve at 1.2\% of the Planck function and a temperature of $\sim$10800 K.
%\item The largest observed enhancements in the NUV clearly lie above this curve and thus require a different model.
\item In the NUV range, the IRIS observations are consistent with the concept of hydrogen Balmer continuum
emission calculated with the 1D static flare models of \citet{ricchiazzicanfield1983}.
\item In the optical and IR regions, RC models do not reproduce the observed enhancements. Only models with photospheric temperature increases, such as in the FLA atmosphere, or empirical input atmospheres modeled with RH, are able to reproduce the CE in these wavelength regions. The emission is then due to H$^-$ and is also  consistent with our data. This indicates that both photospheric and chromospheric emission contribute to the continuum radiation during flares.
\item There are two times where the IRIS slit and the HXR emission are co-temporal and co-spatial. Within the alignment accuracy ($\sim$ 2\arcsec), the maximum RHESSI HXR emission agrees with the maximum NUV CE. Assuming that the NUV continuum arises from recombination of the ionized (due to heating or beam electrons) chromosphere, this requires an instant re-radiation of the deposited energy, with a maximum delay of about 15 seconds.
%Earlier observations often showed a spatial and temporal correlation between the visible continuum and HXR emission  \citep{kruckeretal2011}. 
\item The energy deposited by the non-thermal electrons above 20 keV is at least four times higher than the energy emitted in the continuum. The energy deposited by the non-thermal electrons above $\approx$40 keV matches the energy emitted in the continuum.
\end{enumerate}

Even though earlier observations contained more spectral points, to our knowledge they never sampled such a wide spectral range. For example, \citet{machadorust1974} observed the spectral region from 3530 - 5895 \AA\ during a very strong flare and concluded that their observed enhancement was formed in the chromosphere by free-bound emission of hydrogen at $T$ $\approx$ 8500 K. Recent measurements analyzed white light flares observed with the R,G,B filters on Hinode and found temperatures around 5000-6000 K if they assumed blackbody radiation, and temperatures from 5500-25000 K for hydrogen recombination emission from an optically thin slab \citep{kerrfletcher2014,watanabeetal2013}. However, this result may be influenced by the availability of only three points that are relatively close in wavelength. Observations of the Sun-as-a-star find a WLF CE consistent with a blackbody of about 9000 K  and that the continuum is a large fraction (70\%) of the total radiated energy \citep{kretzschmar2011}. But these values are only reached when subtracting the pre-flare optical continuum before fitting a blackbody, thus assuming an optically thin continuum. If we subtracted pre-flare values in our case, the result would be similar. However, from our modeling, we know that this assumption is not justified, at least in the visible. By combining a large number of observations, including the region around Lyman $\alpha$, \citet{Milliganetal2014} could account for 15 \% of the nonthermal input energy being radiated away in spectral lines and some continua. If we combine their results with our findings, as we now add a point in the UV and better constrain the visible/IR continuum, we are still not seeing $\sim$2/3 of the input energy.  It is possible that it goes into heating, plasma motion, or is radiated away in another spectral region or line.

Different flares have different properties, as is evident with the Balmer jump only sometimes being visible \citep[e.g.,][and references therein]{neidigwiborg1984,neidig1989}, but it is difficult to detect because of various blending effects \citep[e.g.,][]{Kowalskietal2015}. The presence of a CE during
the impulsive phase of this X-class flare was found in IRIS NUV spectra from space for the first time and was interpreted as the Balmer continuum \citep{heinzelkleint2014}. Here we again show that the large observed enhancements very likely are optically thin, cannot be described by blackbody radiation, and require modeling of the Balmer continuum. Specific flare codes like Flarix \citep{Varadyetal2010} and RADYN \citep{allredetal2005} will be used as the next step in the future to improve the realism of the model atmospheres.

%For stellar flares, in particular for flares on M dwarf stars, a Balmer continuum was found for each of the 20 observed flares \citep{kowalskithesis}. They interpreted their observations, which often ranged throughout the whole optical wavelengths above $\sim$ 3400 \AA, with a blackbody component around 10000-12000 K and a Balmer contribution.
%They also found that the Balmer continuum contribution was less during impulsive flares and higher during slow, gradual flares.

The observed co-spatial time delay of the NUV CE behind the HXR of the order of no more than 15 s is compatible with a scenario of heating by electron beams. As the strong NUV CE probably arises from hydrogen recombination in the chromosphere, this also does not pose any problems in terms of penetration depth of the electron beams.

There still are open questions. For example, it is desirable to observe the whole continuum from UV to IR during flares with a low spectral resolution spectrograph, but still with high enough spatial resolution to be able to study spatial variations of the continuum radiation and its relation to HXR sources. This would allow more detailed comparisons with atmospheric models, for example to determine whether the Paschen jump is ever visible and thus to constrain the photospheric temperature increases.

%\section{todo}

%compare to woods et al 2004,2006 who concluded that half of flare energy is radiated away below 190 nm

%cite xu et al 2005 for delays in emission after hxr (in their case integrated over the whole flare)

%color-code locations of large enhancements and oplot HXR path. Any enhancements where no RHESSI?
%mg wing may be an influence

%ext 1-126 is 8542, total
%

\acknowledgments
This work was supported by a Marie Curie Fellowship and the NASA grants NNX13AI63G / NNX14AQ31G. We thank J. P. Wuelser for providing us with the absolute calibration of IRIS and its temporal evolution. We are very grateful to S. Couvidat for creating \textit{hmi.Ic\_45s\_nrt} data for this flare and to P. Scherrer for his explanations of the details of HMI. We thank A. Sainz Dalda for comments on the manuscript and J. Ka\v{s}parov\'{a} for her help with adapting the MALI code to account for non-thermal hydrogen collisional rates. We acknowledge helpful discussions that took place at ISSI.
SK acknowledges the support by the Swiss National Science Foundation (200021-140308), by the European Commission through HESPE (FP7-SPACE-2010-263086), and through NASA contract NAS 5-98033 for RHESSI.  PH was supported by the FP7 project F-CHROMA under the EC contract no.\,606862 and by the
project RVO:67985815.

\bibliographystyle{apj}
\bibliography{journals,ibisflare}

%% for next paper %%%%%%%%%%%%%%%%%
%http://adsabs.harvard.edu/abs/2014A%26A...561A..98S, Sasso: rising filament in flare region
%Most of the profiles in the filament show very broad Stokes I absorptions and complex and spatially variable Stokes V signatures. The inversion of the profiles revealed evidence of multiple unresolved atmospheric blue- and redshifted components of the He i lines within a sin- gle resolution element (? 1 arcsec), with supersonic velocities of up to ? 110 km s?1.

%Thus, Kuckein et al. (2009) and Xu et al. (2012) studied the vector magnetic field of ac- tive region filaments by analysing spectropolarimetric data in the He i 10830  lines, finding the highest field strengths measured in filaments so far, around 600-700 G

\end{document}